\documentclass[11pt]{article}
\usepackage{amsmath}
\usepackage{amssymb}
\usepackage{hyperref}
\usepackage{cancel}
\usepackage{graphicx}
\usepackage[normalem]{ulem}
\usepackage{color}

\numberwithin{equation}{section}

\definecolor{blue-violet}{rgb}{0.54, 0.17, 0.89}
\definecolor{PineGreen}{cmyk}{0.92, 0, 0.59, 0.25}
\definecolor{OliveGreen}{cmyk}{0.64, 0, 0.95, 0.40}
\definecolor{RawSienna}{cmyk}{0, 0.72, 1, 0.45}
\definecolor{Gray}{cmyk}{0, 0, 0, 0.50}

\begin{document}

\begin{titlepage}
\vskip 2.0 cm

\begin{center} {\huge{\bf Spontaneous scalarization in (A)dS gravity\\ at zero temperature   \\ \vskip 0.3 cm }}

\vskip 2.5 cm
{\Large{\bf Alessio Marrani$^1$}, {\bf Olivera Miskovic$^2$}, and {\bf Paula  Quezada Leon$^{2}$}}

\vskip 1.0 cm

$^1${\sl Instituto de F\'{i}sica Teorica, Dep.to de F\'{i}sica,\\Universidad de Murcia, Campus de Espinardo, E-30100, Spain}\\
\texttt{jazzphyzz@gmail.com}\\

\vskip 0.5 cm

$^2${\sl Instituto de F\'{\i}sica, Pontificia Universidad Cat\'{o}lica
de Valpara\'{\i}so,\\Casilla 4059, Valpara\'{\i}so, Chile\\
\texttt{olivera.miskovic@pucv.cl}\\\texttt{pquezada.l@gmail.com}}\\

\vskip 0.5 cm

 \end{center}

\vskip 2.0 cm

\begin{abstract}
We study spontaneous scalarization of electrically charged extremal black holes in $D\geq 4$ spacetime dimensions. Such a phenomenon is caused by the symmetry breaking due to quartic interactions of the scalar -- Higgs potential and Stueckelberg interaction with electromagnetic and gravitational fields, characterized by the couplings $a$ and $b$, respectively. We use the entropy representation of the states in the vicinity of the horizon, apply the inverse attractor mechanism for the scalar field, and analyze analytically the thermodynamic stability of the system using the laws of thermodynamics. As a result, we obtain that the scalar field condensates on the horizon only in spacetimes which are asymptotically non-flat, $\Lambda \neq 0$ (dS or AdS), and whose extremal black holes have non-planar horizons $k=\pm 1$, provided that the mass $m$ of the scalar field belongs to a mass interval (area code) different for each set of the boundary conditions specified by $(\Lambda ,k)$. A process of scalarization describes a second order phase transition of the black hole, from the extremal Reissner-Nordstr\"{o}m (A)dS one, to the corresponding extremal hairy one. Furthermore, for the transition to happen, the interaction has to be strong enough, and all physical quantities on the horizon depend at most on the effective Higgs-Stueckelberg interaction $am^2-2b$. Most of our results are general, valid for any parameter and any
spacetime dimension.
\end{abstract}
\vspace{24pt} \end{titlepage}

\newpage \tableofcontents \newpage

\section{Introduction}

The no-hair theorem, holding in Einstein-Maxwell theories \cite{Israel,Carter, Ruffini}, states that a  black hole (BH) in four-dimensional asymptotically flat spacetime is determined in a unique way in terms of three physical parameters, namely by its mass $M$, electric charge $Q$ and angular momentum $J$; in other words, all higher BH multipole moments are determined only by $M$, $Q$ and $J$. More properly, such a theorem should be referred to as ``no-independent-multipole-hair'' theorem: higher gravitational multipole moments -- quadrupole and higher -- (and electromagnetic multipole moments -- dipole and higher) are not independent for Kerr-Newman BHs. Moreover, the expectation of astrophysical BHs being with no electric charge yields to reasonably conjecture that, in presence of any type of matter-energy, the class of Kerr BHs is the end point of gravitational collapse and thus the most realistic class of solutions to Maxwell-Einstein equations, being uniquely characterized by $M$ and $J$, with no hair whatsoever (this is the so-called ``Kerr hypothesis''). Current and future observations \cite{CG} are testing this conjecture, which has so far been confirmed in various ways, e.g., by the motion of stars around the supermassive BH at the center of the Milky Way (2020 Nobel Prize), by the observation of gravitational waves from BH mergers (2017 Nobel Prize), and by the observation of the shadow of the supermassive BH at the center of M87 (EHT collaboration).

However, throughout the years, stationary BH solutions, usually referred to as `hairy BHs', with either new global charges (primary hair) or new non-trivial fields not associated to a Gauss law -- even if not independent
from the standard global charges (secondary hair) -- have been found in a number of contexts, in which one or more assumptions underlying the aforementioned no-hair theorem were violated.

In fact, by allowing for more general --nonminimal-- coupling functions of the scalar fields to gravity and electromagnetic fields in the Lagrangian density, a new interesting phenomenon, dubbed ``spontaneous scalarization'', was observed: namely, the destabilization of scalar-free BH solutions and the arising of scalar hair. This typically occurs in a number of theories characterized by a non-minimal coupling of
the scalar fields themselves, such that, at critical values of the coupling, BHs develop a tachyonic instability, and new branches of spontaneously scalarized BHs arise out. Neutron stars in scalar-tensor models in which scalar fields were coupled to the Ricci curvature have been the first framework in which scalarization was observed \cite{Damour}. Since then,
such a phenomenon has been established to occur in a number of contexts, e.g., when BHs are coupled to non-linear electrodynamics \cite{NL1, NL2}, or
surrounded by non-conformally invariant matter \cite{C1, C2}, or in Einstein-Yang-Mills theory \cite{EYM1, EYM2, EYM3}, Skyrme hairy black holes \cite{Skyrme1, Skyrme2}, and black holes with dilatonic hair \cite{dilaton} (see \cite{Kleihaus:2015iea}, and e.g. \cite{review} for a review on asymptotically flat BHs). The first example of scalarization with a conformally coupled scalar field $\frac{1}{6}\phi ^{2}R$ in four dimensions has been discussed in \cite{Bekenstein:1974sf, BocharovaEtAl}. In general, fast rotation of the black hole  can induce a tachyonic instability when the scalar is suitably coupled to the curvature \cite{Dima:2020yac}.

On the other hand, BH scalarization can also be induced by higher curvature term corrections to Einstein gravity, coupled to the scalar fields. Furthermore, when the cosmological constant is negative, $\Lambda <0$, the Reissner-Nordstr\"{o}m (RN) BH is not unique in four dimensions\footnote{In the following treatment, we will denote by RN$_{\Lambda }$ the Reissner-N\"{o}rdstrom BH in presence of a non-vanishing cosmological constant.}, and there are also other static BHs of Einstein-Maxwell gravity that have no continuous spatial symmetries; their horizons are smooth and topologically spherical, and they form bound states with the AdS soliton possessing an arbitrary multipole structure \cite{Herdeiro:2016plq, Herdeiro:2018wub}. In recent years, a number of studies \cite{Doneva:2017bvd}--\nocite{Silva:2017uqg, GB3, Doneva:2018rou, GB5, Silva:2018qhn, GB6}\cite{GB7} has investigated asymptotically anti-de-Sitter (AdS) spontaneously scalarized BHs, in which scalar fields are non-minimally couples to the Ricci scalar and Gauss-Bonnet term \cite{Brihaye:2019dck}, also in higher dimensions \cite{Astefanesei:2020qxk}.
In particular,  the spontaneous scalarization phenomenon is also present in the spinning black holes in the  Einstein-Gauss-Bonnet-scalar theory   \cite{Hod:2020jjy}.

Either due to the existence of gravitating solitons, or due to the existence of some particular scalar-gravity couplings, there is now tantamount evidence that, notwithstanding scalar fields usually do not enjoy a Gauss-like law and consequently are hard to keep in equilibrium with an event horizon without trivializing, BH solutions with scalar hair, aka hairy BHs, exist nevertheless.

At any rate, in presence of non-linear and/or higher-derivative curvature terms, the equations of motion are hard to be solved in analytical way, and
only numerical solutions are currently available. This has quite recently motivated the study of the dynamics of scalarized BHs in the simpler class of Einstein--Maxwell-scalar theories with non-minimal couplings between the scalar and Maxwell fields \cite{Herdeiro:2018wub}, which allows to analytically solve the equations of motion, describing the scalar flow in an intrinsically non-linear way \cite{Konoplya:2019goy, Hod:2020ljo}. Various
non-minimal coupling functions \cite{cf1, cf2} have been considered, as well as dyons including magnetic charges \cite{magn}, axionic-type couplings \cite{axion}, and massive and self-interacting scalar fields \cite{Ramazanoglu:2016kul}--\nocite{Macedo:2019sem, Zou:2019bpt}\cite{Fernandes:2020gay}. Spontaneous scalarization was also discussed in presence of a positive cosmological constant \cite{positive}, as well.

On the other hand, the scalarization that occurs spontaneously in non-extremal BHs in AdS space is an important phenomenon in the context of the gauge/gravity duality, since its dual is provided by holographic superconductor-like systems at constant temperature \cite{Gubser:2008px,Hartnoll:2008vx,Hartnoll:2008kx}. A zero-temperature superconductor/insulator phase transition has been obtained in \cite{Nishioka:2009zj} through a mapping of the BH to the AdS soliton, such that the scalar field condensation occurs in the
AdS soliton background. In this work, however, we are interested in different settings, namely, in a scalarization of the extremal BH. Such a zero-temperature system has a degenerate ground state with non-vanishing entropy, similarly to spin glasses, which have been of particular interest recently in  condensed matter physics due to their intriguing properties  (2021 Nobel Prize).

Within this venue of investigation, in \cite{Marrani:2017uli} zero temperature phase transitions driven by the electric charge of asymptotically AdS extremal BHs in four space-time dimensions were investigated, by exploiting the complex Stueckelberg scalar field as an order parameter. Moreover, it was analytically shown in \cite{Marrani:2017uli} that the (necessarily massive) scalar field can couple to a RN BH in the extremal limit of vanishing Hawking temperature, but only if the magnetic charge vanishes, and if space-time has a non-zero cosmological constant, and if a non-minimal
coupling to gravity is present; in particular, in such a framework asymptotically AdS$_{4}$ BHs admit only spherical and hyperbolic near-horizon geometries.

The aforementioned BH scalarization consists of a spontaneous generation of a scalar field $\Psi $ (through symmetry breaking) around the horizon of a BH in a generic class of theories, which may be described by the Lagrangian density \footnote{This is used within the study of scalarization \cite{Kleihaus:2015iea,Herdeiro:2018wub}, as well as in the construction of holographic superconductors \cite{Franco:2009yz}.}
\begin{equation}
\mathcal{L}_{\mathrm{scalarization}}=\frac{1}{2\varkappa }\,R-\frac{1}{2}\,\left( D_{\mu }\Psi \right) ^{2}-\frac{1}{2}\,m^{2}\Psi ^{2}+P(\Psi )\,\mathcal{I}(g,A)\,,\quad 2\varkappa =16\pi G_{\mathrm{N}}\,,
\end{equation}
where $P(\Psi )\geq 0$ is at least quadratic in $\Psi ,$ and $\mathcal{I}(g,A)$ is any other function of the matter fields (in our particular case, of the electromagnetic field) and the metric. This interaction between the scalar, gravity and electromagnetic field leads to the tachyonic scalar (with $m_{\mathrm{eff}}^{2}<0$) and therefore to spontaneous scalarization. In our case, the matter field appears in a $U(1)$-invariant combination,
\begin{equation}
\mathcal{I}(g,A,\sigma )=\left( \partial \sigma -A\right) ^{2}\,,
\end{equation}
and, in the non-extremal case, it needs at least the term $P(\Psi )=\Psi^{2} $. On the other hand, for the scalarization to occur, one needs that the scalar equation  $m^{2}\Psi -f\,\frac{\mathrm{d}P}{\mathrm{d}\Psi }=0$ (where $f$ is the $\Psi $-independent function) possesses at least two different solutions in $\Psi $, and this fact hints at potentials of the form $P\sim \Psi ^{n}$ ($n>2$) as natural candidates.

As mentioned above, spontaneous scalarization is usually traced back to a tachyonic instability of the scalar field; however, this may not be necessarily the case. In \cite{Myung:2018vug}, the appearance of a scalarized RN BH in the Einstein-Maxwell-scalar theory is attributed to the Gregory-Laflamme-like, or modal instability \cite{Gregory:1993vy}, in the sense that it is based on the $\ell =0$ mode perturbations for scalar and tensor fields, and not on a tachyonic field; in other words, the linearized scalar equation of motion includes a non-tachyonic effective mass term which develops negative potential near the horizon from the Maxwell kinetic term. In particular, the coupling $\mathrm{e}^{\alpha \phi ^{2}}F^{2}$ has been considered in \cite{Myung:2018vug}, also pointing out that this exponential potential is very similar to $1+\alpha \phi ^{2}$ from the point of view of the bifurcation points in the field equations. The authors performed numerical computations for the scalar field in the background of the RN BH, finding the threshold of the Gregory-Laflamme instability, $\alpha >\alpha _{\mathrm{th}}(q)$.

In \cite{Marrani:2017uli} and in the present paper, we investigate the spontaneous scalarization of extremal BHs. As discussed in \cite{Brihaye:2019kvj} and recalled above, BHs undergo spontaneous scalarization
for sufficiently large scalar-tensor couplings $\gamma $, usually attributed to a tachyonic instability of the scalar field system. In the extremal case,
a new domain for negative $\gamma $ appears, because of the fact that the near-horizon geometry of a nearly extremal charged black hole is AdS$_{2}\times \mathbb{S}^{2}$. When the extremal BH is not asymptotically flat (for instance, in the case when $\Lambda < 0$), then other geometries can replace $\mathbb{S}^{2}$, such as $\mathbb{R}^{2}$ or $\mathcal{H}^{2}$.

In the present paper, which develops and extends the results of \cite{Marrani:2017uli}, we analytically describe the process of scalarization, exploited via a second order phase transition, from the extremal RN (A)dS BH to the corresponding extremal hairy BH. In order to study phase transitions, a usual thermal field theory approach based on a free energy as an Euclidean continuation of the action is not applicable. Instead, the relevant thermodynamic quantity is the entropy and, to compute it, we apply the entropy function formalism \cite{Sen:2005wa, Sen:2005iz}. More explicitly, we are going to discuss thermodynamics of static, electrically charged extremal BHs in Einstein-Maxwell gravity, for any asymptotical behaviour and in presence of a complex scalar field, whose non-minimal couplings are described by the nonlinear Stueckelberg interaction (modifying the kinetic term of the scalar fields) and the Higgs potential. For the scalarization transition to take place, the interactions have to be strong enough; our results show that the scalar field condensates on the unique extremal BH horizon only in presence of a non-vanishing cosmological constant ($\Lambda \neq 0$) and of non-planar near-horizon geometries ($k=\pm 1$), provided that the
(necessarily non-vanishing) mass of the scalar field is specified within a suitable interval depending on the boundary conditions specified by $\Lambda
$ and $k$.\bigskip

The plan of the paper is as follows. 
We introduce the $D$-dimensional Einstein-Maxwell system coupled to a complex scalar field in Sec.~\ref{EMs system}, where we also present a class of non-minimal interactions of interest. We then derive a set of algebraic equations that govern the entropic behaviour of the system in Sec.~\ref{Entropy function} where, at the end, we also point out some general features of our treatment. Next, we proceed to solve the equations without the scalar field in different settings in Sec.~\ref{RN en D}, whereas we discuss non-solvable equations in presence of the scalar field in Sec.~\ref{Scalar hair}, finding the critical points of the possible phase transitions. In Sec.~\ref{Critical exponents}, we determine the critical exponents for these transitions, and find a near-critical expression for the entropy in Secs.~\ref{RNnear} and \ref{5DhairNear}, without and with the scalar field, respectively. Since the equation for the radius of the transversal space strongly depends on the space dimension, in Sec.~\ref{Five-dimensional solution} we focus on the five-dimensional case, providing an interpretation of our results. However, at the end of the section, we highlight several general features valid in any dimension. Finally, we summarize our results and related open questions in Sec.~\ref{Conclusions}.

\section{Einstein-Maxwell-scalar systems with \texorpdfstring{$\Lambda \neq 0$}{L =0}}
\label{EMs system}

We consider Einstein-Maxwell gravity coupled to one complex scalar field in $D$-dimensional spacetime, in presence of an arbitrary cosmological constant $%
\Lambda $.\footnote{\label{othermodels} This action can be generalized by adding a scalar-dependent function in front of the Maxwell term which
frequently occurs in supergravity, e.g., in $\mathcal{N}=4 $ pure supergravity, in $\mathcal{N}=2$ axion-dilaton gravity \cite{Cremmer:1977tt}%
, or in the so-called $\mathcal{N}=2$ $T^3$ model (see e.g. (3.15) of \cite{Fre:2011uy}), but we are interested first in understanding a theory with
minimal ingredients, before embedding it in a more specific supergravity.}
The dynamics is described by the action
\begin{equation}
I=\int \mathrm{d}^{D}x\,\sqrt{-g}\,\mathcal{L}=\int \mathrm{d}^{D}x\sqrt{-g}%
\left( \frac{1}{2\varkappa }\,\left( R-2\Lambda \right) -\frac{1}{4}\,F^{2}+%
\mathcal{L}_{\mathrm{S}}\right) \,,  \label{I}
\end{equation}%
where $R=g^{\mu \nu }R_{\mu \alpha \nu }^{\alpha }$ is the scalar curvature
of the spacetime metric field $g_{\mu \nu }(x)$, $F_{\mu \nu }=\partial
_{\mu }A_{\nu }-\partial _{\nu }A_{\mu }$ is the 2-form field strength
associated to the vector potential $A_{\mu }(x)$ and $\mathcal{L}_{\mathrm{S}%
}$ is the scalar field Lagrangian density. We also denote the strength of
gravitational interaction\footnote{%
For comparison with \cite{Marrani:2017uli}, we have to take $D=4$ and $4\pi
G_{\mathrm{N}}=1$, which corresponds to $\varkappa =2.$} by $\varkappa =8\pi
G_{\mathrm{N}}$.

The scalar field $\hat{\Psi}=\Psi (x)\mathrm{e}^{\mathrm{i}\sigma (x)}$ is
coupled non-minimally to the system through a non-linear Stueckelberg
interaction \cite{Stuckelberg}
\begin{equation}
\mathcal{L}_{\mathrm{S}}=-\frac{1}{2}\,\left( \partial \Psi \right) ^{2}-%
\frac{1}{2}\,P(\Psi )\left( \partial \sigma -A\right) ^{2}-V(\Psi )\,,\quad
P(\Psi )\geq 0\,,  \label{S}
\end{equation}
often used to describe first and second order phase transitions of non-extremal black holes in Einstein-Maxwell-tensor gravity \cite{Hartnoll:2008kx,Franco:2009yz}. The usual minimally coupled scalar
Lagrangian $-\frac{1}{2}\,|\partial \hat{\Psi}-\mathrm{i}A\hat{\Psi}|^{2}$
is reproduced in the special case $P(\Psi )=\Psi ^{2}$. Note that $\Psi :=|%
\hat{\Psi}|\geq 0$ and the field $\sigma (x)$ is defined on a compact domain
due to an identification $\sigma +2n\pi =\sigma $, $n\in \mathbb{Z}$. The
full action is thus invariant under the $U(1)$ gauge symmetry
\begin{equation}
\hat{\Psi}\rightarrow \mathrm{e}^{i\lambda }\hat{\Psi},\qquad A_{\mu
}\rightarrow A_{\mu }+\partial _{\mu }\lambda \,,
\end{equation}%
or equivalently
\begin{equation}
\Psi \rightarrow \Psi \,,\qquad \sigma \rightarrow \sigma +\lambda \,,\qquad
A_{\mu }\rightarrow A_{\mu }+\partial _{\mu }\lambda \,.  \label{U(1)}
\end{equation}

Interestingly, the above scalar Lagrangian (\ref{S}) is a special case of
the $\sigma $-model Lagrangian coupled to the electromagnetic field,
\begin{equation}
\mathcal{L}_{\mathrm{S}}= -\frac{1}{2}\,G_{ij}(\varphi )\,D_{\mu}\varphi
^{i}D^{\mu }\varphi ^{j}-V(\Psi ) \,,
\end{equation}%
with the real scalar fields $\varphi ^{i}=(\varphi ^{1},\varphi
^{2})=\left(\Psi ,\sigma \right) $, covariant derivatives
\begin{eqnarray}
D_{\mu }\varphi ^{1} &=&D_{\mu }\Psi =\partial _{\mu }\Psi \,,  \notag \\
D_{\mu }\varphi ^{2} &=&D_{\mu }\sigma =\partial _{\mu }\sigma -A_{\mu }\,,
\end{eqnarray}
and the metric of the scalar manifold given by
\begin{equation}
G_{ij}=\left(
\begin{array}{cc}
1 & 0 \\
0 & P\left( \Psi \right)%
\end{array}
\right) .
\end{equation}
Only the $U(1)$ covariant component (phase) of the complex coordinate $\hat{%
\Psi}$ of the scalar manifold is gauged by the vector $A_{\mu }$.

On the other hand, in the framework under consideration, the field $\sigma
(x)$ is non-physical because it can be gauged away using the transformation $%
\sigma ^{\prime }=\sigma +\lambda =0$. Thus, we can introduce a change of
variables such that $A_{\mu }$ is replaced by the gauge-invariant field $%
\tilde{A}_{\mu }=A_{\mu }-\partial _{\mu }\sigma $ under transformations (%
\ref{U(1)}), and with the electromagnetic field strength remaining the same,
$\tilde{F}_{\mu \nu }=F_{\mu \nu }$. Then $I_{\mathrm{S}}=\int \mathrm{d}%
^{D}x\sqrt{-g}\mathcal{L}_{\mathrm{S}}$ becomes explicitly $\sigma $%
-independent without loss of generality. This is consistent with the fact
that the field equation $\delta I_{\mathrm{S}}/\delta \sigma =0$ is not
independent and therefore it is redundant.

In fact, the variation of the action (\ref{I}), (\ref{S}) with respect to
the independent fields yields to the following equations of motion,
\begin{eqnarray}
\delta g_{\mu \nu } &:&\qquad E_{\nu }^{\mu }=R_{\nu }^{\mu }-\frac{1}{2}%
\,\delta _{\nu }^{\mu }R-\delta _{\nu }^{\mu }\Lambda -\varkappa T_{\nu
}^{\mu }\,,  \notag \\
\delta A_{\nu } &:&\qquad E^{\nu }=\nabla _{\mu }F^{\mu \nu }+P\left( \Psi
\right) \,\tilde{A}^{\nu }\,,  \notag \\
\delta \Psi &:&\qquad E=\square \Psi -\frac{1}{2}\left( \frac{\mathrm{d}V}{%
\mathrm{d}\Psi }+\frac{\mathrm{d}P}{\mathrm{d}\Psi }\,\tilde{A}^{2}\right)
\,,  \notag \\
\delta \sigma &:&\qquad \nabla _{\mu }\left[ P\left( \Psi \right) \tilde{A}%
^{\mu }\right] =0\,,  \label{full_eom}
\end{eqnarray}%
where the energy-momentum tensor reads
\begin{equation}
T_{\mu \nu }=F_{\nu \alpha }F_{\mu }^{\alpha }-\frac{1}{4}\,g_{\mu \nu
}F^{2}+P\left( \Psi \right) \tilde{A}_{\mu }\tilde{A}_{\nu }-\frac{1}{2}%
\,g_{\mu \nu }\left[ \left( \partial \Psi \right) ^{2}+P\left( \Psi \right)
\tilde{A}^{2}+2V\left( \Psi \right) \right] \,.
\end{equation}%
The last field equation (for $\sigma $) can be obtained from the Maxwell
equation\footnote{%
The same occurs in the $U(1)^{6}$-invariant Maxwell-Einstein-axion-dilaton
theory (bosonic sector of $\mathcal{N}=4$ pure supergravity) mentioned in
Footnote \ref{othermodels}, after taking a $U(1)^{2}$-invariant truncation
yielding the (bosonic sector of) $\mathcal{N}=2$ Maxwell-Einstein-dilaton
gravity (see e.g. Section 6 of the lectures \cite{Ferrara:2008hwa}).} as $%
\nabla _{\mu }E^{\mu }=0$. Henceforth, we will set $\sigma =0$ and drop the
tilde above the Maxwell field.

In what follows, we will focus on particular functional forms of the
potentials $P$ and $V$. In fact, in \cite{Marrani:2017uli} it was shown that
the extremal Reissner-Nordstr\"{o}m (RN) black hole in four spacetime
dimensions, non-minimally coupled (with a quartic Stueckelberg function $P$
and $V=\frac 12 \, m^2 \Psi^2$), can suffer from a thermodynamic instability
leading to a spontaneous scalarization when $\Lambda \neq 0$. Such an
instability can be traced back to a spontaneous symmetry breaking, which
thus in turn motivates the introduction of a Higgs scalar potential $V\neq 0$
known to produce a symmetry breaking, and the subsequent investigation of
whether it also triggers a scalarization. More concretely, we will study the
following two non-minimal couplings,
\begin{eqnarray}
P\left( \Psi \right) &=&\Psi ^{2}+\dfrac{a}{4}\,\Psi ^{4}\geq 0\,,  \notag \\
V(\Psi ) &=&\dfrac{1}{2}\,m^{2}\Psi ^{2}+\dfrac{b}{4}\,\Psi ^{4}\,.
\label{P,V}
\end{eqnarray}
The coupling constants $m$, $a$ and $b$ have natural dimensions (length)$%
^{-1}$, (length)$^{D-2}$ and (length)$^{D-4}$, respectively. Recall that the
dimension of the gravitational constant $\varkappa$ is (length)$^{D-2}$.

The above Stueckelberg function $P$ is such that $a=0$ corresponds to the minimal coupling and admits the RN$_{\Lambda }$ black hole with arbitrary $\Lambda $, whereas turning on a strong enough interaction $a\neq 0$ produces
a black hole instability and the formation of a hairy black hole in $D=4$ in
the non-extremal case \cite{Gubser:2008px,Hartnoll:2008vx,Hartnoll:2008kx}
and in the extremal case only if $\Lambda \neq 0$ \cite{Marrani:2017uli}. On
the other hand, $V$ describes a potential for the massive complex scalar
field $\Psi $, whose non-minimal coupling is characterized by the
interaction parameter $b\neq 0$ with a `Mexican hat' shape, known to give
rise, in the thermal case, to a spontaneous symmetry breaking as well as to
a second order phase transition in field theory.


\section{Entropy function}
\label{Entropy function}

We will now analyze the black hole instability due to the formation of scalar hair in the extremal case, when the geometry of the (unique) event horizon is known.

We start and consider a static, electrically charged, spherically symmetric extremal black hole in $D$ spacetime dimensions, with an event horizon $\mathbb{H}$ placed at a distance $r_{h}$ from the center of the black hole,
whose near-horizon geometry has topology AdS$_{2}\times \Sigma _{k}$ (with
constant curvature $k=0,\pm 1$). The respective radii of the AdS$_{2}$
subspace and of the transversal section $\Sigma _{k}$ are denoted by $v_{1}$
and $v_{2}$ (both real and positive). The (mostly positive) near-horizon
spacetime metric is generically their direct product,
\begin{equation}
\mathbb{H}:\quad \mathrm{d}s^{2}=g_{\mu \nu }\,\mathrm{d}x^{\mu }\mathrm{d}%
x^{\nu }=v_{1}\left( -r^{2}\mathrm{d}t^{2}+\frac{\mathrm{d}r^{2}}{r^{2}}%
\right) +v_{2}\,\mathrm{d}\Omega _{D-2}^{2}\,.  \label{horizon}
\end{equation}%
The radial coordinate $r$ measures the distance from the horizon $r_{h}$.
The geometry of $\Sigma _{k}$ with the local coordinates $y^{m}$ is given by
the line element on the unit constant curvature,
\begin{equation}
\mathrm{d}\Omega _{D-2}^{2}=\gamma _{mn}(y)\,\mathrm{d}y^{m}\mathrm{d}%
y^{n}\,.  \label{Omega-D-2}
\end{equation}%
Static and spherically symmetric electromagnetic and scalar fields have on
the horizon the most general form
\begin{equation}
F_{\mu \nu }=\left( \delta _{\mu }^{r}\delta _{\nu }^{t}-\delta _{\nu
}^{r}\delta _{\mu }^{t}\right) \,e\,,\qquad \Psi =u\geq 0\,,  \label{F}
\end{equation}%
where $e$ and $u$ are finite parameters. Therefore, one can choose
\begin{equation}
A_{\mu }=\delta _{\mu }^{t}\,er\,.  \label{A}
\end{equation}%
In $D=4$, of course also a generalized theta term could be added to the
action (\ref{I}); however, as found in \cite{Marrani:2017uli}, an extremal
hole with non-vanishing magnetic charge $p$ does not undergo a spontaneous
scalarization. Thus, we will set $p=0$ throughout.

On the other hand, the boundary conditions are the ones of asymptotically
(A)dS or flat spacetimes, and the electric charge $q$ of the electromagnetic
field is fixed on the boundary. On the horizon, due to the attractor
mechanism \cite{AM}, the scalar field does not depend on its value on the
asymptotic boundary. Thus, one can write down for the scalar field $\Psi $
the attractor boundary conditions on the horizon,
\begin{equation}
\Psi \left( r_{h}\right) =u,\qquad \partial _{\mu }\Psi \left( r_{h}\right)
=0\,.  \label{psi}
\end{equation}

In order to study the stability of this black hole, we have to focus on the
entropy as the relevant thermodynamic quantity, since in the extremal limit
the temperature vanishes and the two event horizons get to coincide. The
above boundary behaviour allows for the entropy function formalism \cite{Sen:2005wa,Sen:2005iz} to be applied for the computation of the entropy.
The procedure starts and define the free energy function, obtained as the
Lagrangian density evaluated on the horizon,
\begin{equation}
f\left( u,v_{1},v_{2},e\right) :=\int\limits_{\mathbb{H}}\mathrm{d}^{D-2}y%
\sqrt{-g}\mathcal{L\,},
\end{equation}%
and its Legendre transform, named entropy function, as
\begin{equation}
\mathcal{E}\left( u,v_{1},v_{2},e\right) :=2\pi \left[ eq-f\left(
u,v_{1},v_{2},e\right) \right] \,,  \label{defE}
\end{equation}%
with $q$ being the (asymptotic, conserved) electric charge. The action
principle restricted to the horizon translates into the fact that the
entropy function is maximized. This enables to find the values of the
near-horizon parameters $u$, $v_{1}$, $v_{2}$, $e$ from the algebraic
equations
\begin{equation}
\frac{\partial \mathcal{E}}{\partial u}=0\,,\qquad \frac{\partial \mathcal{E}%
}{\partial v_{1}}=0\,,\qquad \frac{\partial \mathcal{E}}{\partial v_{2}}%
=0\,,\qquad \frac{\partial \mathcal{E}}{\partial e}=0\,.
\end{equation}%
Therefore, the black hole entropy is the extremum of the entropy function,
\begin{equation}
S=\mathcal{E}_{\mathrm{ex}}\,.  \label{SE}
\end{equation}

In the framework under consideration, one can evaluate the following
quantities in the near-horizon \textit{ansatz} (\ref{horizon}), (\ref{A})
and (\ref{psi}),
\begin{eqnarray}
F^{2} &=&-\frac{2e^{2}}{v_{1}^{2}}\,,\qquad \sqrt{-g}=v_{1}v_{2}^{\frac{D-2}{%
2}}\sqrt{\gamma }\,,  \notag \\
R &=&\frac{k}{v_{2}}\,\left( D-2\right) \left( D-3\right) -\frac{2}{v_{1}}\,,
\end{eqnarray}%
where $\gamma $ is the (positive) determinant of the $\left( D-2\right) $%
-dimensional spacial metric (\ref{Omega-D-2}) of $\Sigma _{k}$. Then, for
arbitrary potentials $P$ and $V$, one finds the free energy function to read
\begin{equation}
f=\Omega _{D-2}v_{1}v_{2}^{\frac{D-2}{2}}\left[ \frac{1}{2\varkappa }\left(
\frac{k}{v_{2}}\left( D-2\right) \left( D-3\right) -\frac{2}{v_{1}}-2\Lambda
\right) +\frac{e^{2}}{2v_{1}^{2}}+\frac{e^{2}}{2v_{1}}\,P(u)-V(u)\right] \,,
\end{equation}%
where the area of the horizon $\mathbb{H}$ is
\begin{equation}
\Omega _{D-2}=\int\limits_{\mathbb{H}}\mathrm{d}^{D-2}y\sqrt{\gamma }\,.
\end{equation}%
It is convenient to introduce the electric charge per surface unit of the
horizon (i.e., the electric charge density) as
\begin{equation}
Q:=\frac{q}{\Omega _{D-2}}\,.  \label{Q}
\end{equation}%
In that way, the entropy function (\ref{defE}) in $D$ dimensions becomes
\begin{eqnarray}
\mathcal{E} &=&2\pi \Omega _{D-2}\left[ Qe-\frac{1}{2\varkappa }\left(
k\left( D-2\right) \left( D-3\right) v_{1}v_{2}^{\frac{D-4}{2}}-2v_{2}^{%
\frac{D-2}{2}}-2v_{1}v_{2}^{\frac{D-2}{2}}\Lambda \right) \right.  \notag \\
&&-\left. v_{2}^{\frac{D-2}{2}}\frac{e^{2}}{2v_{1}}-v_{1}v_{2}^{\frac{D-2}{2}%
}\left( \frac{e^{2}}{2v_{1}}\,P(u)-V(u)\right) \right] \,.
\label{entropy-function}
\end{eqnarray}%
By extremizing the above expression,
\begin{equation}
\dfrac{\partial \mathcal{E}}{\partial u}=0\,,\qquad \frac{\partial \mathcal{E%
}}{\partial v_{1}}=0\,,\qquad \frac{\partial \mathcal{E}}{\partial v_{2}}%
=0\,,\qquad \frac{\partial \mathcal{E}}{\partial e}=0\,,
\end{equation}%
we find the equations for the parameters on the horizon (such that $%
v_{1}v_{2}e\neq 0$),
\begin{equation}
\begin{array}{llll}
0=\dfrac{\partial \mathcal{E}}{\partial u} & \Rightarrow & 0= & 2V^{\prime
}\left( u\right) v_{1}-P^{\prime }\left( u\right) e^{2}\,,\medskip \\
0=\dfrac{\partial \mathcal{E}}{\partial v_{1}} & \Rightarrow & 0= & \dfrac{1%
}{2\varkappa }\left[ k\left( D-2\right) \left( D-3\right) -2\Lambda v_{2}%
\right] -\dfrac{e^{2}v_{2}}{2v_{1}^{2}}-V\left( u\right) v_{2}\,,\medskip \\
0=\dfrac{\partial \mathcal{E}}{\partial v_{2}} & \Rightarrow & 0= & \dfrac{1%
}{2\varkappa }\left[ kv_{1}\left( D-3\right) \left( D-4\right)
-2v_{2}-2\Lambda v_{1}v_{2}\right] \medskip \\
&  &  & +\,\dfrac{e^{2}v_{2}}{2v_{1}}-v_{2}\left[ V\left( u\right) v_{1}-%
\dfrac{1}{2}\,P\left( u\right) e^{2}\right] \,,\medskip \\
0=\dfrac{\partial \mathcal{E}}{\partial e} & \Rightarrow & 0= & Q-\dfrac{e}{%
v_{1}}\,v_{2}^{\frac{D-2}{2}}-P\left( u\right) \,e\,v_{2}^{\frac{D-2}{2}}.%
\end{array}
\label{EOM (V y P)}
\end{equation}%
These algebraic equations are valid for any function $P\left( u\right) $ and
$V\left( u\right) $, in any\footnote{%
The cases $D=1,2,3$ do not yield interesting results in the present
treatment, which thus understands $D\geq 4$.} $D\geq 4$, as long as the
scalar field does not depend on the asymptotic conditions.


\subsection{On the general features of our treatment}
\label{GS}
So far, we have presented a gravitational theory coupled to a scalar field, and we highlighted the method which we will use in order to investigate the properties of extremal black holes. Before getting involved into the details of computations, it is here worth listing some remarks. \smallskip

(\textit{i}) Indirect evidence would suggest that hairy extremal black holes should be generally existing. One crucial piece of evidence is provided by holographic superconductors, which are studied at finite temperature $T$ (as duals to non-extremal black holes coupled to a St\"{u}ckelberg scalar field), but which also admit a well defined near-extremal limit, $T\rightarrow 0$ \cite{Horowitz:2009ij}, which exhibits the breakdown of the formalism based on the Euclidean continuation of the action. Another evidence that extremal black holes could undergo phase transitions is provided by a phenomenon, similar to the Meissner effect, observed at $T=0$ in extremal Kerr and Kerr-Newman black holes embedded in an external magnetic field \cite{Astorino:2015naa,Bicak:2015lxa,Astorino:2016xiy}. Last but not least, spin glasses are QFT systems, dual to $T=0$ extremal black holes via the AdS/CFT correspondence, which have non-vanishing entropy and exhibit phase transitions, notwithstanding the vanishing of the temperature. All the above arguments strongly suggest the existence of extremal black holes with scalar condensates.\smallskip

(\textit{ii}) Black hole scalarization is the condensation of a scalar field in proximity of the event horizon, and it is thus legitimate and appropriate to investigate it by exploiting the entropy function formalism, as it has the great advantage to overcome the aforementioned mathematical issues.\smallskip

(\textit{iii}) We do not discuss the existence nor the global properties of particular black hole solutions. In fact, we touch upon global properties of black holes only as far as the cosmological constant enters the treatment or not, and also when exploiting the entropy function formalism (which is based on the assumption that the electric charge is fixed on the boundary, and the attractor mechanism takes place for the scalar field). If hairy black holes fulfilling these conditions exist (and the point (\textit{i}) suggests a positive answer), then their entropy has to be characterized by the features discussed in this work. \medskip

All in all, our approach does not rely on the specification of a particular extremal black hole solution; this is actually a major advantage, as the corresponding nonlinear differential equations are quite difficult to be solved, especially for $D>4$, which is of utmost interest as far as holographic applications are concerned. Moreover, let us recall that backreaction is inherently taken into account in our treatment, in which the space of parameters ($m$, $\Lambda $, $a$, $b$) is arbitrary, and computations are analytic. One can then conclude that our results should be conceived as a first, necessary step towards the (numerical, or semi-analytic) study of extremal black hole solutions.\medskip

On the other hand, our approach is restricted only to phase transitions between extremal black holes, because the entropy function method relies on the existence of a horizon. It does not include other possible solutions in the theory, such as solitons, which could also be relevant. For instance, in \cite{Horowitz:2010jq}, the authors show that the zero temperature AdS soliton has lower free energy than the extremal black hole in AdS space, thus a phase transition would naturally go in its direction. However, the results of ref.~\cite{Horowitz:2010jq} are not comparable to the ones presented in this work, as they describe a phase transition that occurs in a different thermodynamic ensemble (the fixed chemical potential vs. fixed charge), and it involves only planar black holes, which are excluded from our study.


\section{Extremal \texorpdfstring{RN$_{\Lambda }$}{L} black hole \label{RN en D}}

We start and analyze the case without the scalar field, $u=0$. The solution
is the extremal Reissner-Nordstr\"{o}m black hole in the spacetime with arbitrary cosmological constant $\Lambda $, which we will denote as RN$_{\Lambda }$. We will henceforth choose $P(u)$ and $V(u)$ to be given by (\ref{P,V}), and thus $P(0)=0=V(0)$. The equations (\ref{EOM (V y P)})
describing the extremal RN$_{\Lambda }$ black hole become
\begin{eqnarray}
0 &=&\dfrac{1}{2\varkappa }\left[ \frac{k}{v_{2}}\left( D-2\right) \left(
D-3\right) -2\Lambda \right] -\frac{e^{2}}{2v_{1}^{2}}\,,  \notag \\
0 &=&\dfrac{1}{2\varkappa }\left[ k\left( D-3\right) \left( D-4\right) -%
\frac{2v_{2}}{v_{1}}-2v_{2}\Lambda \right] +\frac{e^{2}v_{2}}{2v_{1}^{2}}\,,
\notag \\
0 &=&Q-\frac{ev_{2}^{\frac{D-2}{2}}}{v_{1}}.  \label{eqs u=0}
\end{eqnarray}%
Such three equations solve three parameters $e,$ $v_{1}$ and $v_{2}$ for the
fixed charge $Q$ (assuming $ev_{1}v_{2}\neq 0$). The extremum of $\mathcal{E}
$ corresponds to the black hole entropy $\mathring{S}(Q):=S_{u=0}(Q)$%
.\medskip

Depending on $\Lambda $ and $k$, we can distinguish the following cases.


\subsection{\texorpdfstring{$\Lambda =0$}{L =0}}

When the spacetime is asymptotically flat, $\Lambda =0$, the solution is the
extremal RN$_{\Lambda =0}$ (which we will simply denote by RN) black hole,
whose horizon is necessarily spherical, $k=1$, and the volume $\Omega
_{D-2}=(D-1)\pi ^{\frac{D-1}{2}}/\Gamma \left( \frac{D+1}{2}\right) $
becomes the area of the surface of the sphere $\mathbb{S}^{D-1}$. The
solutions of the system (\ref{eqs u=0}) read%
\begin{eqnarray}
v_{1} &=&\frac{1}{\left( D-3\right) ^{2}}\left( \frac{\varkappa Q^{2}}{%
\left( D-2\right) \left( D-3\right) }\right) ^{\frac{1}{D-3}},  \notag \\
v_{2} &=&\left( \frac{\varkappa Q^{2}}{\left( D-2\right) \left( D-3\right) }%
\right) ^{\frac{1}{D-3}},  \notag \\
e &=&\frac{Q}{\left( D-3\right) ^{2}}\left( \frac{\varkappa Q^{2}}{\left(
D-2\right) \left( D-3\right) }\right) ^{\frac{4-D}{2(D-3)}}.
\end{eqnarray}%
The entropy of the extremal RN black hole is evaluated as the extremum of
the entropy function (\ref{entropy-function}) with the above parameters,
yielding
\begin{equation}
\mathring{S}_{\Lambda =0}(Q)=\frac{2\pi \Omega _{D-2}}{\varkappa }\left(
\frac{\varkappa Q^{2}}{\left( D-2\right) \left( D-3\right) }\right) ^{\frac{%
D-2}{2\left( D-3\right) }}\,.  \label{SRN}
\end{equation}%
We conclude that the entropy $\mathring{S}(q)$ behaves as $q^{\frac{D-2}{D-3}%
}$, or $qe$, as expected. In $D=4$, one obtains $\Omega _{2}=4\pi $ and $%
\mathring{S}_{\Lambda =0}=\pi \Omega _{2}Q^{2}$, so that we reproduce the
known result $\mathring{S}_{\Lambda =0}=q^{2}/4$ \cite{Sen:2007qy}.


\subsection{\texorpdfstring{$k =0$}{k=0}}

A planar event horizon ($k=0$) is supported only by $\Lambda <0$, in which
case the equations (\ref{eqs u=0}) lead to the solution
\begin{equation}
v_{1}=-\frac{1}{2\Lambda }\,,\qquad v_{2}=\left( -\frac{\varkappa Q^{2}}{%
2\Lambda }\right) ^{\frac{1}{D-2}},\qquad e=\sqrt{-\frac{1}{2\varkappa
\Lambda }}\,,
\end{equation}%
and (\ref{entropy-function}) yields the entropy
\begin{equation}
\mathring{S}_{k=0}(Q)=\frac{2\pi \Omega _{D-2}Q}{\sqrt{-2\varkappa \Lambda }}%
\,.  \label{Splanar}
\end{equation}
In $D=4$, the formulae for $\mathring{S}_{\Lambda =0}(Q)$ (\ref{SRN}) and $%
\mathring{S}_{k=0}(Q)$ (\ref{Splanar}) allow to retrieve the known results
\cite{Marrani:2017uli}.


\subsection{\texorpdfstring{$\Lambda \neq 0$, $k\neq 0$}{L =0, k=0} \label%
{RN-Lambda-k}}

Next, we can focus onto the spacetimes with $\Lambda \neq 0$, as well as on
spherical or hyperbolic horizons ($k\neq 0$). The last equation of the
system (\ref{eqs u=0}) is solved by%
\begin{equation}
e=Qv_{1}v_{2}^{\frac{2-D}{2}}\,,  \label{eq1}
\end{equation}%
from which the middle equation of (\ref{eqs u=0}) becomes%
\begin{equation}
k\left( D-3\right) \left( D-4\right) v_{1}-2v_{2}-2v_{2}v_{1}\Lambda +\frac{%
\varkappa Q^{2}v_{1}}{v_{2}^{D-3}}=0\,,  \label{eq3}
\end{equation}%
yielding the solution
\begin{equation}
v_{1}=\frac{2v_{2}^{D-2}}{k\left( D-3\right) \left( D-4\right)
v_{2}^{D-3}-2\Lambda v_{2}^{D-2}+\varkappa Q^{2}}\,.  \label{v1 (v2)}
\end{equation}%
In turn, this implies from (\ref{eq1}) that%
\begin{equation}
e=\frac{2Qv_{2}^{\frac{D-2}{2}}}{k\left( D-3\right) \left( D-4\right)
v_{2}^{D-3}-2\Lambda v_{2}^{D-2}+\varkappa Q^{2}}\,.  \label{e(v1_v2)}
\end{equation}%
All parameters are now expressed in terms of $v_{2}$, which, by virtue of
the first equation of (\ref{eqs u=0}), satisfies
\begin{equation}
2\Lambda v_2^{D-2}-k(D-2)(D-3) v_2^{D-3}+\varkappa Q^2=0\,.  \label{eq2}
\end{equation}
This inhomogeneous polynomial equation in $v_{2}$ has degree $D-2$, thus it
can be solved analytically only in special cases, namely $D=4$, $5$ and $6$,
for which (\ref{eq2}) becomes
\begin{eqnarray}
D &=&4:\qquad \Lambda v_{2}^{2}-kv_{2}+\frac{\varkappa Q^{2}}{2}=0\,,  \notag
\\
D &=&5:\qquad \Lambda v_{2}^{3}-3kv_{2}^{2}+\frac{\varkappa Q^{2}}{2}=0\,,
\notag \\
D &=&6:\qquad \Lambda v_{2}^{4}-6kv_{2}^{3}+\frac{\varkappa Q^{2}}{2}=0\,.
\label{D=4,5,6}
\end{eqnarray}
The case $D=4$ has been discussed in \cite{Marrani:2017uli} in full
generality, so in the present paper we will consider the case $D=5$ in
detail.

The $v_{2}$-dependent expression of the entropy $\mathring{S}(Q)$, obtained
from (\ref{entropy-function}), reads
\begin{equation}
\mathring{S}(Q)=\frac{2\pi \Omega _{D-2}}{\varkappa }\,v_{2}^{\frac{D-2}{2}%
}\geq 0\,.  \label{S without hair}
\end{equation}
For $D=4$, this result matches the one in \cite{Marrani:2017uli}, where $%
v_{2}^{k}=\frac{1}{2\Lambda }\left( k-\sqrt{1-2\varkappa \Lambda Q^{2}}%
\right) $. For $D=5$, one obtains
\begin{equation}
D=5:\qquad \mathring{S}(Q)=\frac{2\pi \Omega _{3}}{\varkappa }\,v_{2}^{\frac{%
3}{2}}\,,
\end{equation}
where $v_{2}(Q)$ has to be determined from (\ref{eq2}) with $D=5$. Explicit
solutions of $v_{2}(Q)$ and the corresponding black hole entropy will be
discussed in the treatment below.

It is here worth remarking that the inhomogeneous polynomial of degree $D-2 \geq 2$ in $v_{2}$ given by \eqref{eq2} is characterized by the discriminant $\delta_{D}$,
defined in terms of a determinant of the $(2D-5) \times (2D-5)$ Sylvester matrix of this polynomial and its first derivative, or in terms of the resultant of the polynomial and its derivative. Using the properties of the resultant, we can arrive to the general formula
\begin{equation}
\delta _{D}=(-1)^{\frac{(D-2)(D-3)}{2}}(\varkappa
Q^{2})^{D-4}(D-2) ^{D-2}\left[ -(D-3)^{2D-5}k^{D-2}+\left(
2\Lambda \right) ^{D-3}\varkappa Q^{2}\right] .
\end{equation}
For $D=4,5,6$, respectively, it reads
\begin{eqnarray}
\delta _{4} &=&4\left( k^{2}-2\varkappa \Lambda
Q^{2}\right) \,,  \notag \\
\delta _{5} &=&108\,\varkappa Q^{2}\left(
8k^{3}-\varkappa Q^{2}\,\Lambda ^{2}\right) \,,  \notag \\
\delta _{6} &=&256\varkappa ^{2}Q^{4}\left( 8\varkappa
\Lambda ^{3}Q^{2}-3^{7}k^{4}\right) \,.  \label{Dis}
\end{eqnarray}
The relation between $Q$, $\Lambda $ and $k$ (where $k=\pm 1$), through the
discriminant, determines the corresponding space of the extremal RN$%
_{\Lambda }$ black hole solutions. In $D=4$ and $D=5$, there are examples in
which the vanishing discriminant does not admit a critical point at which
the phase transition occurs, and thus the point is stable; for example, in $%
D=4$ without the scalar hair, this corresponds to the relation\footnote{%
\label{BPS} When embedded in (gauged) $\mathcal{N}=2$ supergravity, we
expect these particular sectors in the space of parameters to be related to
BPS states, which are stable solutions. For example, the relation of the
type $\delta _{D}(k,\Lambda ,\varkappa Q^{2})=0$ appears in Eq.~(371) of
\cite{Gallerati:2019mzs}, and it has the form $\Gamma ^{m}\theta _{m}=k$,
which holds for the $\frac{1}{4}$-BPS solutions (as it is known for
asymptotically AdS$_{4}$ extremal black holes), where $\Gamma ^{i}$ denotes
the electric-magnetic charges (in our case, $i=1$ and $\Gamma ^{i}\sim \sqrt{%
\varkappa Q^{2}}$), and $\theta _{i}\sim \Lambda $ is related to the gauging
of the theory itself.} $2\varkappa \Lambda Q^{2}=1$ \cite{Marrani:2017uli}.


\section{Scalar hair}
\label{Scalar hair}

In this Section we will allow for the existence of a nontrivial scalar
field, $u>0$, for arbitrary couplings $a$, $b$ and $m$ in the functions $P$
and $V$ given by (\ref{P,V}), and we will study the conditions of its
existence near the horizon of the corresponding extremal black hole. In this
framework, the system (\ref{EOM (V y P)}) becomes
\begin{eqnarray}
0 &=&\left( m^{2}+bu^{2}\right) v_{1}-\left( 1+\frac{a}{2}\,u^{2}\right)
e^{2}\,,  \notag \\
0 &=&\dfrac{1}{\varkappa }\left[ \rule{0pt}{13pt}k\left( D-2\right) \left(
D-3\right) -2\Lambda v_{2}\right] -\dfrac{e^{2}v_{2}}{v_{1}^{2}}-\left(
m^{2}u^{2}+\dfrac{b}{2}\,u^{4}\right) v_{2}\,,  \notag \\
0 &=&\dfrac{1}{\varkappa }\left[ k\left( D-3\right) \left( D-4\right) -\frac{%
2v_{2}}{v_{1}}-2\Lambda v_{2}\right] +\dfrac{e^{2}v_{2}}{v_{1}^{2}}-v_{2}%
\left[ \left( m^{2}u^{2}+\dfrac{b}{2}u^{4}\right) -\left( u^{2}+\frac{a}{4}%
u^{4}\right) \frac{e^{2}}{v_{1}}\right] \,,  \notag \\
0 &=&Q-\left( \dfrac{1}{v_{1}}+u^{2}+\frac{a}{4}u^{4}\,\right) e\,v_{2}^{%
\frac{D-2}{2}}\,.  \label{EOM ab}
\end{eqnarray}%
For a non-vanishing scalar hair, $u$ can be obtained from the first
equation,
\begin{equation}
u=\sqrt{\frac{e^{2}-m^{2}v_{1}}{bv_{1}-\frac{1}{2}\,ae^{2}}}\,,
\label{sol u}
\end{equation}%
and it is well defined if $2bv_{1}-ae^{2}\neq 0$ and the radicand is
positive.
The above equations are highly nonlinear and
not possible to solve analytically for arbitrary $m$, $\Lambda$, $k$, $a$,
$b$, even in the simplest four-dimensional case. This is why we will focus
our attention to the behaviour of these equations in the vicinity of the
critical point, where the scalar field is either zero or very small.
Consistently, the previous treatment of the RN$_{\Lambda }$ extremal black
hole can be retrieved from the limit $u\rightarrow 0^{+}$ of (\ref{sol u}).
The resolution of the second, third and fourth equations of (\ref{EOM ab})
allows to determine $e$, $v_{1}$ and $v_{2}$ in terms of the non-vanishing
scalar hair $u$. As far as we know, this would provide the first example of
the so-called `inverse attractor mechanism' (cfr.e.g. \cite{Saraikin-Vafa})
for asymptotically non-flat ($\Lambda \neq 0$) extremal black holes.


\subsection{Critical limit}

By taking the `critical limit' $u\rightarrow 0^{+}$ within the assumption $%
2bv_{1}-ae^{2}\neq 0$, when $m^{2}>0$, Eq.~(\ref{sol u}) allows to determine
the following relation among the critical values of the horizon parameters :
\begin{equation}
e_{c}^{2}=m^{2}v_{1c}\,.  \label{e crit}
\end{equation}%
In turn, this allows to rewrite the second, third and fourth equation of the
system (\ref{EOM ab}) respectively as follows :%
\begin{eqnarray}
\,k\left( D-2\right) \left( D-3\right) &=&\left( 2\Lambda +\dfrac{\varkappa
e_{c}^{2}}{v_{1c}^{2}}\right) v_{2c}\,,  \notag \\
k\left( D-3\right) \left( D-4\right) &=&\left( 2\Lambda +\frac{2}{v_{1c}}-%
\frac{\varkappa e_{c}^{2}}{v_{1c}^{2}}\right) v_{2c}\,,  \notag \\
Q_{c} &=&\dfrac{e_{c}}{v_{1c}}\,v_{2c}^{\frac{D-2}{2}}.  \label{Eq_c}
\end{eqnarray}


\subsubsection{\texorpdfstring{$\Lambda =0$}{L=0}}

When $\Lambda =0$ and thus $k=1$, a solution exists only for a critical
value of the scalar mass,
\begin{equation}
m_{c}^{2}=\frac{D-2}{\varkappa \left( D-3\right) }\,,
\end{equation}
and for any value of the black hole electric charge density $Q$, since
\begin{eqnarray}
v_{1c} &=&\frac{1}{\left( D-3\right) ^{2}}\left( \frac{\varkappa Q^{2}}{%
\left( D-2\right) \left( D-3\right) }\right) ^{\frac{1}{D-3}}\,,\qquad
v_{2c}=\frac{\varkappa Q^{2}}{\left( D-2\right) \left( D-3\right) }\,,
\notag \\
e_{c} &=&\pm \frac{Q\,}{\left( D-3\right) ^{2}}\left( \frac{\varkappa Q^{2}}{%
\left( D-2\right) \left( D-3\right) }\right) ^{\frac{4-D}{2(D-3)}}\,.
\end{eqnarray}%
Since $Q$ remains arbitrary, such a solution does not correspond to an
isolated critical point for the black hole, but it corresponds rather to a
`critical line' : $u=0$ and $u\neq 0$ can in principle co-exist for any $Q$.


\subsubsection{\texorpdfstring{$k =0$}{k=0}}

When $k=0$ and thus $\Lambda <0$, again the solution exists only for a
critical value of the scalar mass,
\begin{equation}
m_{c}^{2}=\frac{1}{\varkappa }\,,
\end{equation}%
and the corresponding `critical' line is described by the parameters
\begin{equation}
v_{1c}=-\frac{1}{2\Lambda }\,,\qquad v_{2c}=\left( -\frac{\varkappa Q^{2}}{%
2\Lambda }\right) ^{\frac{1}{D-2}}\,,\qquad e_{c}=\pm \sqrt{-\frac{1}{%
2\varkappa \Lambda }}\,.
\end{equation}


\subsubsection{\texorpdfstring{$\Lambda \neq 0$, $k\neq 0$}{L=0, k=0}\label%
{Lambda-k}}

We leave the study of `critical lines' for further future investigation.
Instead, we focus on the cases with $D\geq 4$, $\Lambda \neq 0$, $k\neq 0$,
and $\varkappa m\neq 1$. Then, the solution for the critical point is
unique,
\begin{equation}
v_{1c}=\frac{\Delta }{2\Lambda }\,,\qquad v_{2c}=\frac{k\left( D-3\right)
\Delta }{2\Lambda \left( \varkappa m^{2}-1\right) }\,,\qquad e_{c}^{2}=\frac{%
m^{2}\Delta }{2\Lambda }\,,  \label{uno}
\end{equation}
with
\begin{equation}
\Delta :=\varkappa m^{2}\left( D-3\right) -\left( D-2\right) \,.  \label{due}
\end{equation}
The critical charge reads
\begin{equation}
Q_{c}=\sqrt{\frac{2\Lambda m^{2}}{\Delta }}\,\left( \frac{k\left( D-3\right)
\Delta }{2\Lambda \left( \varkappa m^{2}-1\right) }\right) ^{\frac{D-2}{2}},
\label{Qc}
\end{equation}%
where we chose $Q_{c}>0$ for simplicity's sake\footnote{%
This does not imply any loss of generality, because $Q\rightarrow -Q$ only
corresponds to the change of direction of the electric field ($e\rightarrow
-e$), as it can be seen from (\ref{F}).}. In $D=4$, (\ref{Qc}) returns the
known result.

From (\ref{uno}), the positivity of the parameters $v_{1c}$, $v_{2c}$ and $%
e_{c}^{2}$ requires
\begin{equation}
m^{2}>0\,,\qquad \Delta \Lambda >0\,,\qquad k\left( \varkappa m^{2}-1\right)
>0\,.  \label{positivity}
\end{equation}
For example, the scalar mass can take the value $%
\varkappa m^{2}=2$ only in $D\geq 5$, for which $\Delta =D-4>0$, and then
necessarily $\Lambda >0$ in order to have real $e_{c}$.
Similarly, each mass of the scalar field different than
$0$, $1/\varkappa$ and $(D-2)/(D-3)\varkappa$, uniquely determines the signs
of $k$ and $\Lambda$, and possible dimensions with non-trivial solutions for
the near-horizon parameters. In the same line of reasoning, in general, solving the inequalities (\ref{positivity}) determines the following three
regions in the space of parameters,
\begin{equation}
\begin{array}{llll}
\text{\textbf{A}.} & \Lambda >0\,,\quad (\Delta >0)\,, & k=1\,, & \dfrac{D-2%
}{D-3}<\varkappa m^{2}\,,\smallskip \\
\text{\textbf{B}.} & \Lambda <0\,,\quad (\Delta <0)\,, & k=1\,, &
1<\varkappa m^{2}<\dfrac{D-2}{D-3}\,,\smallskip \\
\text{\textbf{C}.} & \Lambda <0\,,\quad (\Delta <0)\,,\quad & k=-1\,,\quad &
0<\varkappa m^{2}<1\,.%
\end{array}
\label{A-B-C}
\end{equation}%
It is here worth remarking that (\ref{A-B-C}) defines \textit{area codes}
(in the sense of \textit{non-uniqueness} of the consistent solutions, as
discussed in \cite{Moore} and \cite{Giry}) for the inverse attractor
mechanism in asymptotically non-flat extremal black holes in $D\geq 4$
spacetime dimensions; such area codes are defined only by the scalar mass $%
m^{2}$ for one given background determined by $\Lambda $ and $k$.

Using (\ref{S without hair}) and (\ref{uno}), one can compute the critical
entropy as%
\begin{equation}
S_{c}=\mathring{S}(Q_{c})=\frac{2\pi \Omega _{D-2}}{\varkappa }\left( \frac{%
k\left( D-3\right) \Delta }{2\Lambda \left( \varkappa m^{2}-1\right) }%
\right) ^{\frac{D-2}{2}}\,.  \label{Sc}
\end{equation}%
This expression coincides with the one of \cite{Marrani:2017uli} when $D=4$.


\section{Critical exponents}
\label{Critical exponents}

In order to determine the near-critical behavior of the horizon parameters $u $, $v_{1}$, $v_{2}$, $e$ and $Q$, we expand them around the critical point
:
\begin{eqnarray}
u &=&\left( A\epsilon \right) ^{n}+\left( K\epsilon \right) ^{z}+\cdots \,,
\notag \\
v_{1} &=&v_{1c}+B\epsilon ^{w}+U\epsilon ^{h}+\cdots \,,  \notag \\
v_{2} &=&v_{2c}+C\epsilon ^{p}+N\epsilon ^{c}+\cdots \,,  \notag \\
e &=&e_{c}+E\epsilon ^{s}+R\epsilon ^{d}+\cdots \,,  \notag \\
Q &=&Q_{c}+\epsilon \,,  \label{expansion}
\end{eqnarray}%
where a small value of $\epsilon $ can be either positive or negative, with
the sign determined by the dynamics. First, we will have to find the
critical exponents of the leading order $n$, $w$, $p$, $s$, and then the
ones of the sub-leading order, namely $z$, $h$, $c$, $d$. Note that $u=%
\tilde{A}\epsilon ^{n}+\tilde{K}\epsilon ^{z}$, with\ $\tilde{A}\epsilon
^{n}>0$, has been conveniently written using $\tilde{A}=A^{n}$ and $\tilde{K}%
=K^{z}$.

By definition of critical point, by plugging the near-critical expansion (%
\ref{expansion}) into the field equations (\ref{EOM ab}), the finite orders
cancel out, and at the leading order in $\epsilon $ the system of equations
of motion can be written (after using (\ref{e crit})) in the following
matrix form :
\begin{equation}
\left(
\begin{array}{cccc}
\left( b-\frac{am^{2}}{2}\right) v_{1c} & m^{2} & 0 & -2e_{c} \\
\frac{m^{2}v_{2c}}{2} & -\frac{m^{2}v_{2c}}{v_{1c}^{2}} & \dfrac{\Lambda }{%
\varkappa }+\frac{m^{2}}{2v_{1c}} & \frac{e_{c}v_{2c}}{v_{1c}^{2}} \\
0 & \frac{\left( 1-\varkappa m^{2}\right) v_{2c}}{\varkappa v_{1c}^{2}} &
\frac{\varkappa m^{2}-2}{2\varkappa v_{1c}}-\frac{\Lambda }{\varkappa } &
\frac{e_{c}v_{2c}}{v_{1c}^{2}} \\
e_{c}\,v_{2c}^{\frac{D-2}{2}} & -\frac{e_{c}}{v_{1c}^{2}}\,v_{2c}^{\frac{D-2%
}{2}} & \dfrac{\left( D-2\right) e_{c}}{2v_{1c}}\,v_{2c}^{\frac{D-4}{2}} &
\dfrac{v_{2c}^{\frac{D-2}{2}}}{v_{1c}}%
\end{array}%
\right) \left(
\begin{array}{c}
\rule{0pt}{20pt}\left( A\epsilon \right) ^{2n} \\
\rule{0pt}{20pt}B\epsilon ^{w} \\
\rule{0pt}{20pt}C\epsilon ^{p} \\
\rule{0pt}{20pt}E\epsilon ^{s}%
\end{array}%
\right) =\left(
\begin{array}{c}
\rule{0pt}{20pt}0 \\
\rule{0pt}{20pt}0 \\
\rule{0pt}{20pt}0 \\
\rule{0pt}{20pt}\epsilon%
\end{array}%
\right) .
\end{equation}%
In order to have a non-vanishing solution in terms of $(A^{2n},B,C,E)$ when $%
\epsilon \to 0$, all leading critical exponents must necessarily be equal,
namely
\begin{equation}
2n=w=p=s=1\,.  \label{lead}
\end{equation}%
Similarly, the next-to-leading critical exponents are found to read
\begin{equation}
z=\frac{3}{2}\,,\qquad h=c=d=2\,.  \label{sublead}
\end{equation}


\section{\label{RN-Lambda critical} Extremal \texorpdfstring{RN$_{\Lambda }$}{RNL} black hole near criticality}
\label{RNnear}

Without scalar hair (i.e., with $u=0$), the treatment of Sec.~\ref{RN-Lambda-k} yields that a consistent solution exists for any $Q$. If $Q$ is near its the critical value, one can then expand $\mathring{S}(Q)=
\mathring{S}(Q_{c}+\epsilon )$ near $Q_{c}$.

Firstly, it is convenient to solve $v_{1}(Q)$ near $Q_{c}$ from Eq.~(\ref{eq2}) by exploiting the method of successive approximations. From (\ref{expansion}), (\ref{lead}) and (\ref{sublead}), it holds that
\begin{eqnarray}
Q &=&Q_{c}+\epsilon \,,  \notag \\
v_{2}(Q) &=&v_{2c}+\mathring{C}\epsilon +\mathring{N}\epsilon ^{2}+\mathcal{O%
}(\epsilon ^{3})\,.
\end{eqnarray}
For $D\geq 4$ and $k\neq 0$, the critical coefficients at orders $\epsilon $
and $\epsilon ^{2}$ are respectively found to be
\begin{eqnarray}
\mathring{C} &=&\frac{\varkappa }{D-2}\sqrt{\frac{2m^{2}\Delta }{\Lambda }}%
\left( \frac{k\left( D-3\right) \Delta }{2\Lambda \left( \varkappa
m^{2}-1\right) }\right) ^{\frac{4-D}{2}}\,,  \notag \\
\mathring{N} &=&\frac{k\varkappa \left( \varkappa m^{2}-1\right) }{\left(
D-3\right) \left( D-2\right) ^{2}}\left( \frac{k\left( D-3\right) \Delta }{%
2\Lambda \left( \varkappa m^{2}-1\right) }\right) ^{4-D}\left[ \rule%
{0pt}{14pt}D-2+2\left( D-2+\Delta \right) \left( \varkappa m^{2}-2\right) %
\right] \,.  \notag \\
&&
\end{eqnarray}
Thus, the entropy $\mathring{S}$, given by \eqref{S without hair} of the
corresponding extremal RN$_{\Lambda }$ black hole, is
\begin{equation}
\mathring{S}=\mathring{S}(Q_{c}+\epsilon )=S_{c}+\epsilon \,\mathring{S}%
_{1}+\epsilon^{2}\mathring{S}_{2}+\mathcal{O}(\epsilon ^{3})\,,
\end{equation}
where the critical value $S_{c}$ is given by (\ref{Sc}), and the first two
critical coefficients read\footnote{%
Note that Eqs.~(\ref{S10,S20}) match the results obtained in $D=4$ in \cite%
{Marrani:2017uli}, in which a direct expansion of $v_{2}(Q)$ yields $%
\mathring{C}=2\sqrt{\frac{m^{2}\left( m^{2}-1\right) }{\Lambda }}$ and $%
\mathring{N}=\left( 2m^{2}-1\right) ^{3}$.}
\begin{eqnarray}
\mathring{S}_{1} &=&S_{c}\,\frac{D-2}{2v_{2c}}\,\mathring{C}=2\pi
\Omega_{D-2}\,e_{c}\,,  \notag \\
\mathring{S}_{2} &=&S_{c}\,\frac{D-2}{2v_{2c}}\left( \mathring{N}+\frac{D-4}{%
4v_{2c}}\,\mathring{C}^{2}\right) =\frac{\pi \Omega _{D-2}}{2\Lambda m^{2}}%
\,v_{2c}^{\frac{2-D}{2}}\left[ \left( 2\varkappa m^{2}-3\right) m^{2}\Delta
+2\Lambda \mathbf{c}_{0}\right] \,,  \label{S10,S20}
\end{eqnarray}
where we have introduced the constant
\begin{equation}
\mathbf{c}_{0}:=\frac{m^{2}\Delta }{2\Lambda \left( D-2\right) }\,\left[\rule%
{0pt}{14pt}4\left( D-2\right) -\varkappa m^{2}\left( 3D-8-2\Delta \right) %
\right] \,.  \label{c0}
\end{equation}
In what follows, we will see that the value of $\mathbf{c}_0$ determines a
strong interaction regime, the only possible regime where the black hole
scalarization might occur.


\section{\label{D=5 SS}Scalar hair and spontaneous scalarization}
\label{5DhairNear}

Now, we compute the black hole entropy in presence of non-vanishing scalar
hair, i.e. with $u\neq 0$. To this end, we have to solve the equations of
motion (\ref{EOM ab}) in the proximity of the critical point by exploiting
the near-critical expansion (\ref{expansion})--(\ref{sublead}) : this will
allows us to determine the leading order coefficients ($A$, $B$, $C$, $E$)
as well as the sub-leading order coefficients ($K$, $U$, $N$, $R$), which
for $u\neq 0$, from (\ref{expansion}) with (\ref{lead}) and (\ref{sublead}),
enter the near-critical expansions
\begin{eqnarray}
u &=&\sqrt{A\epsilon }+\sqrt{K^{3}\epsilon ^{3}}+\mathcal{O}(\epsilon
^{5/2})\,,  \notag \\
v_{1} &=&v_{1c}+B\epsilon +U\epsilon ^{2}+\mathcal{O}(\epsilon ^{3})\,,
\notag \\
v_{2} &=&v_{2c}+C\epsilon +N\epsilon ^{2}+\mathcal{O}(\epsilon ^{3})\,,
\notag \\
e &=&e_{c}+E\epsilon +R\epsilon ^{2}+\mathcal{O}(\epsilon ^{3})\,.
\label{parameters}
\end{eqnarray}%
By recalling that we use the explicit expressions of $P(u)$ and $V(u)$ given
by (\ref{P,V}), we interestingly find that the solution depends only on the
effective parameter
\begin{equation}
\mathbf{c}:=m^{2}a-2b\,,  \label{c}
\end{equation}%
and not on the single parameters $m^{2}$, $a$ and $b$. In fact, we find the
leading order coefficients to be uniquely determined as\footnote{%
For $D=4$, (\ref{ABCE}) reduces to the formulae of \cite{Marrani:2017uli},
in which the expressions for $B$ and $E$ have been interchanged.}%
\begin{eqnarray}
A &=&-\frac{8k\Lambda \left( D-3\right) e_{c}v_{1c}^{2}v_{2c}^{-\frac{D}{2}}%
}{\left( D-2\right) \left( \mathbf{c}-\mathbf{c}_{0}\right) }\,,  \notag \\
B &=&\frac{2\varkappa k\left( D-3\right) e_{c}v_{1c}^{2}v_{2c}^{-\frac{D}{2}}%
}{\left( D-2\right) \left( \mathbf{c}-\mathbf{c}_{0}\right) }\,\left[ \rule%
{0pt}{14pt}\mathbf{c}\left( D-3\right) -e_{c}^{2}\,\left( D-4\right) \right]
\,,  \notag \\
C &=&\,\frac{2\varkappa e_{c}v_{2c}^{\frac{4-D}{2}}}{\left( D-2\right)
\left( \mathbf{c}-\mathbf{c}_{0}\right) }\,\left[ \rule{0pt}{14pt}\mathbf{c}%
+e_{c}^{2}\left( \varkappa m^{2}-2\right) \right] \,,  \notag \\
E &=&\frac{k(D-3)v_{2c}^{-\frac{D}{2}}}{\left( D-2\right) \left( \mathbf{c}-%
\mathbf{c}_{0}\right) }\,\left[ \rule{0pt}{14pt}\mathbf{c}\left(
D-2+4\Lambda v_{1c}\right) -\frac{\varkappa e_{c}^{4}}{v_{1c}}\,\left(
D-4\right) \right] ,  \label{ABCE}
\end{eqnarray}%
where $\mathbf{c}_{0}$ is defined by (\ref{c0}). Without the interaction
(i.e., for $\mathbf{c}=0$), the above critical coefficients do not vanish,
even though we would expect $u=0$ in such a case; as we will find below,
this can be traced back to the fact that, in general, the entropy or its
derivatives change in a non-continuous way at the critical point, hinting at
a possible scalarization.

Eqs.~(\ref{ABCE}) imply some useful relations, which allows to express $C$
and $E$ only in terms of $A$ and $B$,
\begin{eqnarray}
C &=&\frac{kv_{2c}^{2}}{\left( D-3\right) ^{2}v_{1c}^{2}}\left( B-\frac{%
\varkappa e_{c}^{2}v_{1c}}{2}\,A\right) \,,  \notag \\
E &=&\frac{e_{c}}{2v_{1c}}\,B-\frac{v_{1c}}{4e_{c}}\,\mathbf{c}A\,,
\label{E}
\end{eqnarray}
and in turn $B$ in terms of $A$ only,
\begin{equation}
B=\frac{\varkappa A}{4\Lambda }\,\left[ \rule{0pt}{14pt}\left( D-4\right)
e_{c}^{2}-\left( D-3\right) \mathbf{c}\right] \,.  \label{B}
\end{equation}%
Thus, all coefficients $B$, $C$ and $E$ can be written in terms of $A$ only.

Again, the inequalities (\ref{positivity}) guarantee that all coefficients
are real and well defined; moreover, we also need $\mathbf{c}\neq \mathbf{c}%
_{0}$. We observe that the very existence of the scalar hair $u\neq 0$
determines the sign of the coefficient $A$ itself; for the time being, we
notice that, when $k\neq 0$, $\Lambda \neq 0$ and $\mathbf{c}\neq \mathbf{c}%
_{0}$, it holds that%
\begin{equation}
\mathrm{sgn}\left( A\right) =\mathrm{sgn}\left( k\Lambda \left( \mathbf{c}%
_{0}-\mathbf{c}\right) \right) \,.  \label{sgnA}
\end{equation}

The sub-leading order critical coefficients $K$, $U$, $N$, $R$ are given by
cumbersome expressions which are not very illuminating, and they will not be
explicitly written here, especially because the knowledge only of (\ref{ABCE}%
) suffices to determine $S$, which turns out not to depend on higher-order
coefficients, up to the order $\epsilon ^{3}$. In fact, from (\ref{EOM (V y
P)}), by exploiting $\dfrac{\partial \mathcal{E}}{\partial v_{1}}=0$ and $%
\dfrac{\partial \mathcal{E}}{\partial e}=0$ only, the entropy (\ref%
{entropy-function}) is given by a simple expression,%
\begin{equation}
S=2\pi \Omega _{D-2}\left( \frac{1}{\varkappa }+\frac{e^{2}}{2}\,P\right)
v_{2}^{\frac{D-2}{2}},
\end{equation}%
which matches $\mathring{S}$ when $u=0\Rightarrow P(0)=0$; cfr. Eq.~(\ref{S
without hair}). Again, near the critical point, the entropy behaves as%
\begin{equation}
S=S_{c}+\epsilon \,S_{1}+\epsilon ^{2}S_{2}+\mathcal{O}(\epsilon ^{3})\,,
\label{S-crit}
\end{equation}%
where the critical value $S_{c}$ is given by (\ref{Sc}). The next-to-leading
term reads
\begin{equation}
S_{1}=S_{c}\,\left( \frac{D-2}{2v_{2c}}\,C+\frac{\varkappa e_{c}^{2}}{2}%
\,A\right) \,,
\end{equation}%
which in turn, by means of the identity $\mathring{C}-C=\frac{\varkappa
e_{c}^{2}v_{2c}}{D-2}\,A$, simplifies down to%
\begin{equation}
S_{1}=S_{c}\,\frac{D-2}{2v_{2c}}\,\mathring{C}=\pi \Omega _{D-2}\sqrt{\frac{%
2m^{2}\Delta }{\Lambda }}.
\end{equation}%
This yields an interaction-free expression, namely,
\begin{equation}
S_{1}=\mathring{S}_{1}=2\pi \Omega _{D-2}\,e_{c}\,,  \label{leading-S}
\end{equation}%
which is indeed satisfied according to Eq.~(\ref{S10,S20}) : up to the order
$\mathcal{O}(\epsilon ^{2})$, the entropy is continuous, $S(Q)=S_{c}+\left(
Q-Q_{c}\right) \,\mathring{S}_{1}+\mathcal{O}(\left( Q-Q_{c}\right) ^{2})$,
since the presence of a non-vanishing scalar hair $u\neq 0$ does \textit{not}
change it. Thus, a phase transition, if any at all is involved in the
scalarization, cannot be of the first order.

Without using the equations of motion, but rather only (\ref{Eq_c}) and $%
k\neq 0$, the next-to-next-to-leading entropic term\ $S_{2}$ (i.e., the one
multiplying the quadratic contribution in $\epsilon $ in (\ref{S-crit})) can
be computed to read
\begin{eqnarray}
S_{2} &=&2\pi \Omega _{D-2}E +S_{c}\left[ -\frac{\varkappa c\Delta }{%
16\Lambda }\,A^{2}+\frac{\varkappa m^{2}}{2}\,AB-\frac{2\varkappa
m^{2}\Lambda ^{2}}{\Delta ^{2}}\,B^{2}-\frac{\varkappa \Lambda }{\Delta }%
\,E^{2}\right.  \notag \\
&&+\frac{\Lambda ^{2}(D-2)(\varkappa m^{2}-1)^{2}}{(D-3)^{2}\Delta ^{2}}
\left( \frac{(D-4)(\varkappa m^{2}-1)}{2}\,C^{2}+2k\,BC\right)  \label{S2b}
\\
&&+\left. \varkappa \sqrt{\frac{m^{2}\Delta }{2\Lambda }}\left( -AE+\frac{%
4\Lambda ^{2}}{\Delta ^{2}}\,BE-\frac{2k(D-2)(\varkappa m^{2}-1)\Lambda ^{2}
}{\Delta ^{2}(D-3)}\,CE\right) \right] ,  \notag
\end{eqnarray}
where the coefficients $A$, $B$, $C$ and $E$ are given by the first Eq.~(\ref%
{ABCE}) and (\ref{E})--(\ref{B}). As anticipated, (\ref{S2b}) result does
\textit{not} depend on the sub-leading order critical coefficients $K$, $U$,
$N$, $R$.

According to the first law of thermodynamics, the system will undergo a
phase transition with the newly formed scalar hair $u\neq 0$ only if the
corresponding black hole entropy near the critical point is larger than the
entropy in absence of scalar hair (i.e., for $u=0$). Thus, by virtue of the
interaction free result (\ref{leading-S}) a phase transition will take place
if
\begin{equation}
S_{2}>\mathring{S}_{2};
\end{equation}
since the formula (\ref{S2b}) depends on many free parameters such as $m^{2}$%
, $\Lambda $, $k$, $\mathbf{c}$, it is difficult to draw immediate
conclusions for a generic $D>3$.

For what concerns the role of the coupling constants $a$ and $b$ occurring
in $P$ and $V$ as given by (\ref{P,V}), we recall that the former modifies
the kinetic term of the scalar field whereas the latter modifies the
potential. As observed above, the effective dynamics near the critical point
only depends on the effective combination $\mathbf{c}$ defined by (\ref{c}),
explicitly depending on the scalar mass. Nevertheless, from the treatment of
spontaneous scalarization given below for $D=5$, we will see that this is
just a near-critical feature, and it does not characterize the whole horizon
dynamics.


\section{Spontaneous scalarization in \texorpdfstring{$D=5$}{D=5}}
\label{Five-dimensional solution}

In order to obtain explicit results, we will now focus on $D=5$. The case $D=4$ has been studied in detail in \cite{Marrani:2017uli}, and the treatment
is valid for any $a$ and $b$ by the replacement $a\rightarrow a-\frac{2b}{m^{2}}$. In this Section we will analyze in detail the case $\Lambda \neq 0$
and $k=\pm 1$, as discussed, in absence as well as in presence of scalar
hair, respectively in Sections \ref{RN-Lambda-k} and \ref{Lambda-k}.


\subsection{Extremal \texorpdfstring{RN$_{\Lambda }$}{RNL} black hole}

Without scalar hair (i.e., for $u=0$), the extremal black hole is the
electrically charged RN$_{\Lambda }$ one\footnote{%
For $k=1$, in $D=5$ the solution is also named Tangherlini black hole, with
near-horizon geometry $AdS_{2}\times S^{3}$ \cite{Tang1, Tang2}.}. For $D=5$%
, Eq.~(\ref{D=4,5,6}) has three solutions,
\begin{eqnarray}
v_{2}^{(1)} &=&\frac{1}{\Lambda }\left( k+\frac{Y}{2}+\frac{2}{Y}\right) \,,
\notag \\
v_{2}^{(2)} &=&\frac{1}{\Lambda }\left[ k-\frac{Y}{4}-\frac{1}{Y}+\frac{%
\mathrm{i}\sqrt{3}}{2}\left( \frac{Y}{2}-\frac{2}{Y}\right) \right] \,,
\notag \\
v_{2}^{(3)} &=&\frac{1}{\Lambda }\left[ k-\frac{Y}{4}-\frac{1}{Y}-\frac{%
\mathrm{i}\sqrt{3}}{2}\left( \frac{Y}{2}-\frac{2}{Y}\right) \right] \,,
\label{v}
\end{eqnarray}
where
\begin{equation}
Y:=\left( 8k-2\varkappa Q^{2}\Lambda ^{2}+2Q\Lambda \sqrt{-\frac{4\delta
_{5} }{27Q^{2}}}\right) ^{1/3}\,.  \label{Y}
\end{equation}
The number of real solutions depends on the discriminant $\delta _{5}$ given
by (\ref{Dis}); depending on its sign, we can distinguish three cases :

\begin{figure}[tbp]
\centering
\includegraphics[width=.4\textwidth]{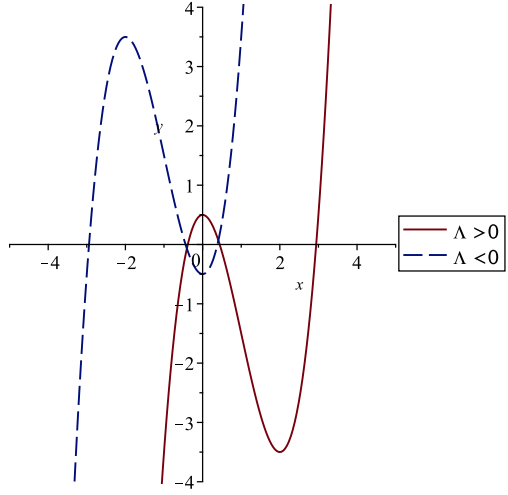}
\caption{{\protect\small The graphs show the zeroes of the polynomial $F(x)$
with the positive discriminant, $\mathbf{Q} ^{2}<\protect\lambda ^{2}$,
corresponding to the solutions for horizon radius $x=v_2 >0$ of the
three-dimensional space $\Sigma_k$. When $\Lambda>0$ (continuous line),
there are two solutions. When $\Lambda<0$ (dashed line), there is only one
solution. The number of the solutions does not depend of the values of the
parameters; the graphs are drawn for $\mathbf{Q}=1$ and $\Lambda=\pm 2$.}}
\label{graphF}
\end{figure}

\begin{enumerate}
\item $\delta _{5}>0$. There are three different real roots because the
cubic function in $v_{2}$ has a threefold intersection with the $v_{2}$%
-axis. This holds for
\begin{equation}
\delta _{5}>0\quad \Rightarrow \quad Q^{2}<\frac{8k}{\varkappa \Lambda ^{2}}
\,,  \label{Q small}
\end{equation}
which in turn is possible only for spherical horizons ($k=1$), and when the
charge (and the mass) of the extremal RN$_{\Lambda }$ black hole are small.
By simplifying the notation as $v_{2}=:x$, $\Lambda ^{-1}=:\frac{\lambda }{2}
$ and $\frac{\varkappa Q^{2}}{2}=:\mathbf{Q}^{2}$, we can analyze the curve $%
F(x)=x^{3}-\frac{3\lambda }{2}\, x^{2}+\frac{\lambda }{2}\,\mathbf{Q}^{2}$.
The condition of positive discriminant means that $\mathbf{Q} ^{2}<\lambda
^{2}$. It holds that lim$_{x\rightarrow \pm \infty }F(x)=\pm \infty $,
whereas the extrema are at $0=F^{\prime }(x)=3x\left( x-\lambda \right) $,
namely at the points $x=0$ and $x=\lambda $, for which $F(0)= \frac{\lambda
}{2}\mathbf{Q}^{2}$ resp. $F(\lambda )=\frac{\lambda }{2} \left( \mathbf{Q}%
^{2}-\lambda ^{2}\right) $. Since $F^{\prime \prime }(x)=3\left(2x-\lambda
\right) $, it then holds that $F^{\prime \prime }(0)=-3\lambda $ and $%
F^{\prime \prime }(\lambda )=3\lambda $. Therefore, there is always one
minimum and one maximum, depending on the sign of $\lambda $. When $\lambda
>0$, both extrema are in the $x\geq 0$ half-plane, namely at $x=0$ and $%
x=\lambda $. Since $F^{\prime \prime }(0)=-3\lambda $, $x=0$ is the maximum
which corresponds to $F_{\max }=\frac{\lambda }{2}\mathbf{Q}^{2}>0$;
analogously, since $F^{\prime \prime }(\lambda )=3\lambda >0$, $x=\lambda $
is the minimum, corresponding to $F_{\min }=\frac{\lambda }{2}\left( \mathbf{%
Q}^{2}-\lambda ^{2}\right) <0$. This shape of the cubic curve $F(x)$ is such
that it intersects the $x$-axis exactly twice in the positive half-plane,
and this implies that there are always, for any choice of parameters,
exactly two positive real zeros (with the third real zero being always
negative, and thence to be disregarded). Therefore, for $\lambda >0$, there
are always two different solutions $v_{2}$, one in the interval $0<v_{2}<%
\frac{2}{\Lambda }$ (corresponding to $0<x<\lambda $) and another in the
interval $v_{2}>\frac{2}{\Lambda }$ (corresponding to $x>\lambda $). When $%
\lambda <0$, both extrema are in the $x\leq 0$ half-plane, namely at $x=0$
and $x=\lambda $. Since $F^{\prime \prime }(0)=-3\lambda $, $x=0$ is the
minimum, corresponding to $F_{\min }=\frac{\lambda }{2}\mathbf{Q}^{2}<0$,
while, since $F^{\prime \prime }(\lambda )=3\lambda $, $x=\lambda $ is the
maximum, corresponding to $F_{\max }=\frac{\lambda }{2}\left( \mathbf{Q}%
^{2}-\lambda ^{2}\right) >0$. This shape of the cubic curve $F(x)$ is such
that it intersects the $x$-axis exactly once in the positive half-plane, and
this implies that there is always, for any choice of parameters, exactly one
positive real zero (with the other two real zeros being always negative, and
thus to be disregarded). Therefore, for $\lambda <0$, there is always
exactly one solution $v_{2}>0$. The function $F(x)$ in the cases of positive
and negative cosmological constant is shown in Fig.~\ref{graphF}, where only
its right half-plane zeros correspond to the physical solutions of the
radius $v_2$.

\item $\delta _{5}=0$. There are two different real roots $v_{2}^{(1)}$ and $%
v_{2}^{(2)}$ in (\ref{v}), because one of them is multiple ($%
v_{2}^{(2)}=v_{2}^{(3)}$) as the cubic function intersects only once the $%
v_{2}$-axis, and a second time it touches it. This means that the charge is
fixed, and the positivity of $Q^{2}$ allows only for $k=1$,
\begin{equation}
\delta _{5}=0\quad \Rightarrow \quad Q^{2}=\frac{8}{\varkappa \Lambda ^{2}}%
\,,\quad k=1\,.  \label{pre-Q bound}
\end{equation}

\item $\delta _{5}<0$. There is only one real root (the other two being
complex), and thus%
\begin{equation}
\delta _{5}<0\quad \Rightarrow \quad Q^{2}>\frac{8k}{\varkappa \Lambda ^{2}}%
\,.  \label{Q bound}
\end{equation}
As we can see, this is possible for any kind of non-planar horizon (i.e.,
for $k=\pm 1$) : hyperbolic horizons can have arbitrary charge, whereas
spherical horizons must have large charge (and large mass).
\end{enumerate}

However, also in the cases 2 and 3 only positive values $v_{2}^{(i)}$ are
physically meaningful, so only a subset of the solutions corresponding to (%
\ref{pre-Q bound}) and (\ref{Q bound}) will give rise to a consistent radius
of the three-dimensional transversal section $\Sigma _{k}$ of the extremal RN%
$_{\Lambda }$ black hole. The above analysis shows that the number of real
positive solutions for $v_{2}$ is always fixed for given parameters, whose
precise value is not relevant for a qualitative analysis. Below, we list
some explicit examples of the values of these parameters (for simplicity's
sake, we set $\varkappa =1$), corresponding to three broad classes
(distinguished by labels \textbf{A}, \textbf{B} and \textbf{C}, as in (\ref%
{A-B-C})) :

\begin{equation*}
\begin{tabular}{llllll}
$1:\delta _{5}>0\quad \medskip $ & Class \textbf{A}$\quad $ & $\Lambda =1$ &
$k=1$ & $Q=2$ & $v_{2}=1\,,$ $v_{2}=1+\sqrt{3}\,.$ \\
$\medskip $ & Class \textbf{B} & $\Lambda =-1\quad $ & $k=1$ & $Q=2$ & $%
v_{2}=\sqrt{3}-1\,.$ \\
$\medskip $ & Class \textbf{C} & $\Lambda =-1$ & $k=-1\quad $ & ----- & -----
\\
$2:\delta _{5}=0\medskip $ & Class \textbf{A} & $\Lambda =1$ & $k=1$ & $Q=2%
\sqrt{2}$ & $v_{2}=2\,.$ \\
$\medskip $ & Class \textbf{B} & $\Lambda =-1$ & $k=1$ & $Q=2\sqrt{2}$ & $%
v_{2}=1\,.$ \\
$\medskip $ & Class \textbf{C} & $\Lambda =-1$ & $k=-1$ & ----- & ----- \\
$3:\delta _{5}<0\medskip $ & Class \textbf{A} & $\Lambda =1$ & $k=1$ & -----
& ----- {\small (see below, also for $k=-1$)} \\
& Class \textbf{B} & $\Lambda =-1$ & $k=1$ & $Q=3$ & $v_{2}\approx 1.1\,,$
\\
& Class \textbf{C} & $\Lambda =-1$ & $k=-1$ & $Q=3$ & $v_{2}\approx 1.1\,.$%
\end{tabular}%
\end{equation*}
The straight line means that no real positive solution exists for $v_{2}$.
With the above reasoning, we have proved that there is no solution when $%
\delta _{5}\geq 0$ and $k=-1$. Now, we will show that there is no positive
solution for $v_{2}$ when $\delta _{5}<0$ and $\Lambda >0$, for any kind of
non-planar horizon (i.e., for $k=\pm 1$). We start by observing that $\delta
_{5}<0$ implies $Y$ in (\ref{Y}) to be real, and thus $v_{2}^{(2,3)}$ are
complex. Moreover, by setting $k=1$, the first of Eqs.~(\ref{v}) yields
\begin{equation}
v_{2}^{(1)}=1+\frac{Y}{2}+\frac{2}{Y}\,,  \label{v2-k=1}
\end{equation}
where, since $Y$ (\ref{Y}) depends on $\Lambda $ through the product $%
Q\Lambda $ and $\Lambda >0$, we have chosen $\Lambda =1$, obtaining
\begin{equation}
Y=\left( 8-2Q^{2}+2Q\sqrt{Q^{2}-8}\right) ^{1/3}<0\,,\qquad \text{ when }
Q^{2}>8\,,
\end{equation}
where we have assumed $Q>0$ without loss of generality\footnote{%
Indeed, the polynomials (\ref{D=4,5,6}) depend on $Q^{2}$ only.}. Therefore,
for $v_{2}^{(1)}$ in (\ref{v2-k=1}) to be positive, it must hold that $\frac{%
Y}{2}+\frac{2}{Y}>-1$, which is impossible for $Y<0$. On the other hand,
when $k=-1$, the first of Eqs.~(\ref{v}) yields
\begin{equation}
v_{2}^{(1)}=-1+\frac{Y}{2}+\frac{2}{Y}\,,  \label{v2-k=-1}
\end{equation}
where, again, we have chosen $\Lambda =1$, obtaining%
\begin{equation}
Y=\left( -8-2Q^{2}+2Q\sqrt{Q^{2}-8}\right) ^{1/3}<0\,,\qquad \text{when }%
Q^{2}>8\,,
\end{equation}
where again we have assumed $Q>0$ without loss of generality. Therefore, for
$v_{2}^{(1)}$ in (\ref{v2-k=-1}) to be positive, it must hold that $\frac{Y}{%
2}+\frac{2}{Y}>1$, which is impossible for $Y<0$. This means that there is
no positive solution for $v_{2}$, \textit{q.e.d.}.


\subsection{Entropy : with or without hair}

Let us now study the $D=5$ extremal black hole near the critical point $%
Q_{c} $. The near-critical expansions of the extremal black hole entropies,
in absence and presence of a non-trivial scalar hair (corresponding to the RN%
$_{\Lambda }$ resp. the hairy black hole) read
\begin{eqnarray}
\mathring{S} &=&S_{c}+2\pi \Omega _{3}e_{c}\,\epsilon +\mathring{S}%
_{2}\,\epsilon ^{2}+\mathcal{O}(\epsilon ^{3})\,,  \label{S-without} \\
S &=&S_{c}+2\pi \Omega _{3}e_{c}\,\epsilon +\mathring{S}_{2}\,\omega
\epsilon ^{2}+\mathcal{O}(\epsilon ^{3})\,,  \label{S-with}
\end{eqnarray}%
where the quadratic order $\mathring{S}_{2}$ of the entropy $\mathring{S}$
(i.e., with no scalar hair : $u=0$) is given by Eqs.~(\ref{S10,S20}) and (%
\ref{c0}) with $D=5$, yielding%
\begin{equation}
\mathring{S}_{2}=\frac{k\pi \Omega _{3}}{6}\sqrt{\frac{k\Lambda \left(
\varkappa m^{2}-1\right) }{2\varkappa m^{2}-3}}\left( \varkappa
m^{2}-1\right) ^{2}\left( 4\varkappa m^{2}-3\right) \,.  \label{S20}
\end{equation}%
On the other hand, by setting $D=5$ and plugging (\ref{ABCE})--(\ref{B})
into (\ref{S2b}), the quadratic order $S_{2}$ of the entropy $S$ (i.e., with
scalar hair : $u\neq 0$) can be computed to read%
\begin{equation}
S_{2}=\frac{k\pi \Omega _{3}}{12\Lambda }\sqrt{\frac{k\Lambda (\varkappa
m^{2}-1)}{2\varkappa m^{2}-3}}\frac{\left( \varkappa m^{2}-1\right) }{\left(
\mathbf{c}_{0}-\mathbf{c}\right) }^{2}\left[ \rule{0pt}{14pt}\varkappa
m^{4}\left( 2\varkappa m^{2}-3\right) -2\Lambda \mathbf{c}\left( 4\varkappa
m^{2}-3\right) \right] \,.  \label{S2}
\end{equation}%
As resulting from (\ref{leading-S}), at the first order in $\epsilon $ there
is no dependence of the RN$_{\Lambda }$ extremal black hole entropy on the
`effective interaction' $\mathbf{c}$ (\ref{c}). On the other hand, the
dependence on $\mathbf{c}$ appears at the second order in $\epsilon $ in the
expansion of the extremal hairy black hole entropy. Furthermore, $\mathbf{c}%
_{0}$ is given by (\ref{c0}) which, for $D=5$, becomes%
\begin{equation}
\mathbf{c}_{0}=\frac{m^{2}}{6\Lambda }\,(2\varkappa m^{2}-3)(4\varkappa
^{2}m^{4}-13\varkappa m^{2}+12)>0\,,  \label{c0-2}
\end{equation}
which is strictly positive as a consequence of the inequalities (\ref%
{positivity}), as well as from the fact that $4y^{2}-13y+12>0$, $\forall y$.

The comparison of the quadratic terms of (\ref{S-without}) and (\ref{S-with}%
), namely of (\ref{S20}) and (\ref{S2}), allows one to define
\begin{equation}
\omega :=\frac{S_{2}}{\mathring{S}_{2}}=1+\frac{2m^{2}(\varkappa
m^{2}-1)(2\varkappa m^{2}-3)^{3}}{3\Lambda (4\varkappa m^{2}-3)\left(
\mathbf{c}-\mathbf{c}_{0}\right) }\,,
\end{equation}%
which is the unique source of dependence of (\ref{S-with}) on $\mathbf{c}$,
at the second order in $\epsilon $ included.

The factor $\omega $ plays the crucial role of determining which entropy is
larger, and therefore it decides which state of the extremal black hole will
be realised for given electric charge density $Q$, depending on the sign $%
\mathring{S}_{2}$ (\ref{S20}):

\begin{itemize}
\item when $\mathring{S}_{2}>0$, the hairy black hole has larger entropy
than the RN$_{\Lambda }$ black hole and thus a spontaneous scalarization
takes place, when $\omega >1$;

\item when $\mathring{S}_{2}<0$, the hairy black hole has larger entropy
than the RN$_{\Lambda }$ black hole and thus a spontaneous scalarization
takes place, when $\omega <1$.
\end{itemize}

A detailed case study analysis yields that $\mathring{S}_{2}$ (\ref{S20}) is
always positive, so the larger entropy solution always has $\omega >1$,
except in the (example pertaining to the) class \textbf{C} above, in the
interval of scalar masses $\varkappa m^{2}\in \left( \frac{3}{4},1\right) $
: in such a case, $\mathring{S}_{2}$ (\ref{S20}) becomes negative and the
larger entropy solution has $\omega <1$. In all cases, regardless the sign
of $k=\pm 1$, the spontaneous scalarization of the extremal black hole
occurs if the `effective interaction' $\mathbf{c}$ is positive and strong
enough, namely if
\begin{equation}
\mathbf{c}>\mathbf{c}_{0}>0\,.  \label{cond-c}
\end{equation}

Apart from the sign of $\omega -1$, we should also ensure that the scalar
hair $u$ is well defined, i.e., a real (positive) number; from the first of
the near-critical expansions (\ref{parameters}), this depends on the sign of
the coefficient $A$, whose expression, by recalling (\ref{uno}), (\ref{due})
and the first of Eqs.~(\ref{ABCE}), reads for $D=5$ as follows :
\begin{equation}
A=\frac{2k\Lambda \left( \varkappa m^{2}-1\right) ^{2}\sqrt{2km^{2}\left(
\varkappa m^{2}-1\right) }}{3\left( \mathbf{c}_{0}-\mathbf{c}\right) }\,.
\end{equation}%
From (\ref{positivity}), the critical coefficient $A$ is always real and,
since (\ref{cond-c}) yields $\mathbf{c}_{0}-\mathbf{c}<0$, its sign
coincides with the sign of $-k\Lambda $, as expressed by Eq.~(\ref{sgnA}).
Considering all the (examples pertaining to the) classes \textbf{A}, \textbf{%
B} and \textbf{C} discussed above, we obtain the following near-critical
solutions for the scalar field $u$,
\begin{eqnarray}
u_{\mathbf{A},\mathbf{C}}(Q) &\simeq &\sqrt{A(Q-Q_{c})}\,\mathcal{H}\left(
Q_{c}-Q\right) \,,  \notag \\
u_{\mathbf{B}}(Q) &\simeq &\sqrt{A(Q-Q_{c})}\,\mathcal{H}\left(
Q-Q_{c}\right) \,,  \label{u's}
\end{eqnarray}
where $\mathcal{H}(Q)$ is the Heaviside step function which vanishes for
negative arguments. Thus, it is worth remarking that the scalar field always
condensates on only one side of the critical point $Q=Q_{c}$.

To summarize, for each of the (classes defined by the) boundary conditions
determined by the pair $(\Lambda ,k)$, there is a range of the scalar masses
$m^{2}$, as given in (\ref{A-B-C}), for which the spontaneous scalarization
in $D=5$ spacetime dimensions takes place above or below the critical
electric charge density $Q_{c}$, because the corresponding extremal hairy
black hole becomes thermodynamically more stable than the electrically
charged extremal RN$_{\Lambda }$ black hole. In all cases, the process
depends on the `effective coupling constant' $\mathbf{c}$ defined by (\ref{c}%
) : the value of $\mathbf{c}$ has to be larger than $\mathbf{c}_{0}$ given
by (\ref{c0-2}) for the phase transition to occur, as expressed by the
condition (\ref{cond-c}). Recalling that $\epsilon =Q-Q_{c}$ (cfr. the fifth
Eq.~of (\ref{expansion})), from the near-critical expansions (\ref{S-without}%
)--(\ref{S-with}) it is evident that the phase transition determined by the
spontaneous scalarization of the electrically charged extremal RN$_{\Lambda}
$ black hole is characterized by the discontinuity of the response function $%
S^{\prime \prime }(Q)$,
\begin{equation}
\lim_{Q\to Q_{c}^{+}} S^{\prime \prime }(Q)\neq \lim_{Q\to Q_{c}^{-}}
S^{\prime \prime }(Q),
\end{equation}
thus implying that the phase transition that occurs at $Q=Q_c$ is a second
order one.


\subsection{Stueckelberg \textit{versus} Higgs}

We have observed that, near the critical point, only the effective coupling
constant $\mathbf{c}$, defined by (\ref{c}), matters. This feature does not
hold away from criticality, because the Stueckelberg non-minimal interaction
modifies the kinetic term as in the $\sigma $-model, while the Higgs
interaction modifies only the potential. However, at the (unique) event
horizon of the extremal black hole and near criticality, such two
interacting terms do not play separate roles.


\subsubsection{Symmetry enhancement at the critical point?}

The (apriori unreasonable) effectiveness\footnote{%
The qualitative conclusions of the present discussion do not depend on
spacetime dimension $D\geq 4$.} of $\mathbf{c}$ can be better appreciated by
considering the scalar field effective potential\footnote{%
Not to be confused with the so-called `black hole effective potential'.},
obtained as the total potential of the scalar field that includes
electromagnetic and gravitational interaction \cite{Gubser:2008px},
\begin{equation}
V_{\mathrm{eff}}\left( \Psi \right) :=V\left( \Psi \right) +\frac{1}{2}%
\,P\left( \Psi \right) \mathbf{A}^{2}=\frac{1}{2}\left( m^{2}+\mathbf{A}%
^{2}\right) \Psi ^{2}+\frac{1}{4}\left( \dfrac{a}{2}\,\mathbf{A}^2+b\right)
\Psi ^{4}\,,  \label{Veff}
\end{equation}
where
\begin{equation}
\mathbf{A}^{2}:=g^{\mu \nu }A_{\mu }A_{\nu }\,.  \label{AA}
\end{equation}
The equilibrium points for the scalar field are found from $V_{\mathrm{eff}%
}^{\prime }(\Psi)=0$, solved by
\begin{equation}
\Psi =0\,, \quad \text{and}\qquad \Psi _{0}^{2}=-\frac{\mathbf{A}^2+m^{2}}{%
\dfrac{a}{2}\,\mathbf{A}^{2}+b}\,,  \label{extr}
\end{equation}
respectively, corresponding to the local extrema of the potential,
\begin{equation}
V_{\mathrm{eff}}\left( 0\right) =0\,,\quad\text{and}\qquad V_{\mathrm{eff}%
}\left(\Psi _{0}\right) =-\frac{1}{4}\,\frac{\left( \mathbf{A}%
^{2}+m^{2}\right) ^{2}}{\dfrac{a}{2}\,\mathbf{A}^{2}+b}\,.  \label{Vextr}
\end{equation}
The effective mass
\begin{equation}
m_{\mathrm{eff}}^{2}\left( \Psi \right) :=V_{\mathrm{eff}}^{\prime \prime 2}
+\mathbf{A}^{2}+3\left(\dfrac{a}{2}\, \mathbf{A}^{2}+b\right) \Psi ^{2}\,,
\end{equation}
when non-vanishing, determines the unstable (maximum) or stable (minimum)
nature of the critical points. It is an \textit{on-shell} quantity and it
clearly depends not only on the value of the scalar, but also on the value
of the electromagnetic and gravitational fields through $\mathbf{A}^{2}$ (%
\ref{AA}), and therefore it involves the whole dynamics.

At the horizon $\mathbf{A}^{2}(r_{h})=-\frac{e^{2}}{v_{1}}$, and close to
the critical point $\frac{e^{2}}{v_{1}}=m^{2}+\mathcal{O}(\epsilon )$, one
obtains that $\mathbf{A}^{2}+m^{2}=\mathcal{O}(\epsilon )$ vanishes, and the
effective coupling constant $\mathbf{c}$ (\ref{c}) comes from the $\mathbf{A}%
^2$-interaction,
\begin{equation}
\dfrac{a}{2}\,\mathbf{A}^{2}+b=-\dfrac{m^{2}a-2b}{2}+\mathcal{O}(\epsilon)=:-%
\dfrac{\mathbf{c}}{2}+\mathcal{O}(\epsilon )\,.
\end{equation}
Thus, the effective constant $\mathbf{c}$ arises naturally on the horizon
and near criticality, i.e. when $Q\simeq Q_{c}$, and it is invariant under
the exchange
\begin{equation}
am^{2}\leftrightarrow -2b\,.  \label{mab}
\end{equation}
This symmetry concerns all the horizon parameters up to linear order, as
well as the black hole entropy up to quadratic order. However, the equations
of motion in the near-horizon geometry are not invariant under (\ref{mab}).
The fact that this property does not hold for general $Q$ (and thus away
from $Q_{c}$) can be appreciated by continuing the near-$Q_{c}$ power-series
expansion; by doing so, one would find that all horizon parameters, as well
as the entropy, acquire terms which separately depend on $a$ and $b$, and
which thus cannot be expressed only in terms of $\mathbf{c}$. This hints to
a possible symmetry enhancement at $Q\simeq Q_{c}$ related to scale
invariance and the answer, which can be obtained by analyzing near-horizon
asymptotic isometries of the spacetime, surely deserves further
investigation.


\subsubsection{Effective mass and scalar condensation}

On the other hand, to understand why the scalar field condensates and the
scalarization takes place, we have to find its \textit{on-shell} effective
mass at the extremum points, which respectively reads
\begin{equation}
m_{\mathrm{eff}}^{2}(0)=m^{2}+\mathbf{A}^{2}\,,\qquad m_{\mathrm{eff}%
}^{2}\left( \Psi _{0}\right) =-2\left( \mathbf{A}^{2}+m^{2}\right) \,.
\label{meff}
\end{equation}
Since $m_{\mathrm{eff}}^{2}\left( 0\right) $ and $m_{\mathrm{eff}%
}^{2}\left(\Psi _{0}\right) $ have opposite signs, only one of these two
points is a (stable) minimum for a given set of parameters : it ultimately
depends on the whole dynamics, because the equations of motion for $A_{\mu
}(x)$ and $g_{\mu \nu }(x)$ have to be solved. Furthermore, the sign of $m_{%
\mathrm{eff}}^{2}$ itself also depends on the (squared) mass of the scalar
field. Thus, the existence of another non-vanishing minimum $\Psi _{0}\neq 0$
depends on the full fledged dynamics of the electromagnetic, gravitational
and scalar fields, as well as on the (choice of) values of coupling
constants.

Since we are considering extremal black holes, the attractor mechanism \cite%
{AM} is at work, and it yields to a proper (i.e., stable) attractor value $u$
at the event horizon \textit{if and only if} $u$ is a local minimum, i.e. if
it has a positive effective mass, $m_{\mathrm{eff}}^{2}(u)>0$ (see e.g. \cite%
{Goldstein-et-al}). In the framework under consideration, (\ref{meff})
implies in any $D\geq 4$ that
\begin{equation}
m_{\mathrm{eff}}^{2}(u)=-2\left( m^{2}-\frac{e^{2}}{v_{1}}\right) =-2m_{%
\mathrm{eff}}^{2}(0)\,,  \label{meff_eval}
\end{equation}
and the result in turn depends on the solution for $e$ and $v_{1}$ obtained
from other equations.

Eq.~(\ref{meff_eval}) yields that, if $u=0$ is a proper attractor ($m_{%
\mathrm{eff}}^{2}(0)>0$), then $u\neq 0$ is not a proper attractor, but
rather a repeller, with $m_{\mathrm{eff}}^{2}(u)<0$, and the other way
around: if $u\neq 0$ is a proper attractor, then $u=0$ is a repeller
critical point :

\begin{itemize}
\item when $u=0$, a proper and unique attractor ($m_{\mathrm{eff}}^{2}(0)>0$
), then Higgs potential $\approx $ Stueckelberg interaction, and the scalar
hair $u\neq 0$ corresponds to a repeller critical point, with $m_{\mathrm{eff%
}}^{2}(u)<0$;

\item when the scalar hair $u\neq 0$ corresponds to a proper (i.e. stable)
attractor ($m_{\mathrm{eff}}^{2}(u)>0$), $u=0$ is a repeller critical point (%
$m_{\mathrm{eff}}^{2}(0)<0$) and the contribution of the Higgs and
Stueckelberg interactions to $V_{\mathrm{eff}}$ can be interchanged by
\begin{equation}
2b\leftrightarrow -a\,\frac{e^{2}}{v_{1}}\,,
\end{equation}
which follows from (\ref{mab}) and (\ref{meff_eval}).
\end{itemize}

Clearly, without any interaction ($a=b=0$), it holds that $m_{\mathrm{eff}%
}^{2}=m^{2}\geq 0$, so $u=0$ is the only solution, and no phase transition
takes place. On the other hand, the expansion of the effective mass (\ref%
{meff_eval}) near the critical point $Q=Q_{c}$ yields
\begin{equation}
m_{\mathrm{eff}}^{2}(u)=-\frac{2e_{c}}{v_{1c}^{2}}\,\left(e_{c}\,B-2v_{1c}E%
\right) \left( Q-Q_{c}\right) +\mathcal{O}(\left( Q-Q_{c}\right) ^{2})\,.
\end{equation}
Using (\ref{E}) and (\ref{B}), this expression can be recast, in any
dimension, in the following form :
\begin{equation}
m_{\mathrm{eff}}^{2}(u)=2\mathbf{c}A\left( Q_{c}-Q\right) +\mathcal{O}%
(\left( Q-Q_{c}\right) ^{2})>0\,.
\end{equation}
Thus, the effective mass of the scalar condensate is manifestly always
positive on one side of the critical point, because (\ref{u's}) yields that $%
\mathrm{sgn}(Q-Q_{c})=\mathrm{sgn}(A) $ and $\mathbf{c}$ is positive (cfr. (%
\ref{cond-c})). Crossing the critical point, the solution for $u$, as given
by (\ref{u's}) for the various classes under consideration, vanishes and it
yields $m_{\mathrm{eff}}^{2}(0)=0$, becoming the unique attractor point.

It is worth remarking here that, when $\Lambda <0$, it is known that the
scalar equation in AdS$_{D}$ spacetime admits a stable solution (under
mechanical perturbations) for the scalar masses satisfying the
Breitenlohner-Freedman bound \cite{BF1,BF2}, $m_{\mathrm{eff}}^{2}\ell^2 >-%
\frac{(D-1)^{2}}{4}$, where $\ell=\sqrt{\frac{(D-1)(D-2)}{2|\Lambda|}}$ is
the AdS radius. Since in our case $m_{\mathrm{eff}}^{2}$ is positive, this
bound is always satisfied. It does not mean, however, that the system is
also thermodynamically stable. Indeed, the spontaneous scalarization could
occur in the extremal cases when the particular conditions among the
parameters are met, classified by intervals \textbf{A}, \textbf{B} or
\textbf{C}. To prove that such thermodynamically unstable extremal black
hole solutions exist in the whole space, it is necessary to solve the full
fledged equations of motion.\bigskip

Let us now discuss the above results within the classes \textbf{A}, \textbf{B%
} and \textbf{C} introduced in the previous treatment. We will consider the
entropy density per horizon surface unit; for simplicity's sake, we will set
$\Omega _{3}=1$, $|\Lambda|=1$ and $\varkappa =1$. Also, the mass $m^{2}$
will be fixed, so that we can analyze the effect of interactions through the
effective coupling $\mathbf{c}$, but the obtained results are valid
generically within a given class.

\paragraph{Class A}

Within this class, the extremal black hole without scalar hair is an
electrically charged and asymptotically dS$_{5}$ Tangherlini black hole. The
scalar field has mass $m^{2}=2$ and the `effective coupling' is strong
enough if $\mathbf{c}>\frac{2}{3}$, with respect to the value $\mathbf{c}%
_0=\frac 23$. Hence, the entropy density has the form
\begin{equation}
S(Q)=\left\{
\begin{array}{lll}
2\pi+2\pi \,(Q-Q_c) +\dfrac{5\pi }{6}\,\left( Q-Q_{c}\right) ^{2}+\mathcal{%
\cdots }\,, & Q\geq Q_{c}, & \text{{\small Tangherlini~dS}}_{5}~\text{%
{\small BH};} \\
&  &  \\
2\pi+2\pi \,(Q-Q_c) +\dfrac{5\pi }{6}\dfrac{\left( \mathbf{c}-\frac{2}{5}%
\right) }{\left( \mathbf{c}-\frac{2}{3}\right) }\,\left( Q-Q_{c}\right)
^{2}+\cdots \,, & Q\leq Q_{c}, & \text{{\small sph.~hairy~dS}}_5~\text{%
{\small BH},}%
\end{array}
\right.  \label{1-entropy}
\end{equation}
and the order parameter (i.e., the value of the scalar field at the horizon)
reads
\begin{equation}
u(Q)=\left\{
\begin{array}{ll}
0\,, & Q\geq Q_{c}\,,\medskip \\
\frac{2}{\sqrt{3}}\sqrt{\dfrac{Q_{c}-Q}{\mathbf{c}-\frac{2}{3}}}+\cdots \,,
& Q\leq Q_{c}\,.%
\end{array}
\right.  \label{1-u}
\end{equation}
Thus, for large charges, i.e. for $Q\geq Q_{c}$, the extremal black hole is
an asymptotically dS$_{5}$ Tangherlini black hole, with no scalar hair.
Since $\dfrac{\mathbf{c} -\frac{2}{5}}{\mathbf{c} -\frac{2}{3}}>1$, as the
electric charge density decreases to approach $Q_{c}$ from the right, the
scalar field starts to condensate spontaneously on the horizon, because when
$Q\leq Q_{c}$ the corresponding hairy extremal black hole has a higher
entropy. As resulting from (\ref{1-entropy}) (and as discussed above), the
phase transition corresponding to the scalarization is of second order.

It is also instructive to show that, as we go farther from the critical
point, the system begins to depend on both coupling constants $a$ and $b$
(occurring in (\ref{P,V})) separately, and not only on $\mathbf{c}$. For
example, from the critical expansion (\ref{parameters}) of the attractor
horizon value $u$ of the scalar field, the sub-leading term reads
\begin{equation}
\sqrt{(K\epsilon)^{3}}=\sqrt{\frac{\epsilon ^{3}}{A}}\frac{-33\mathbf{c}%
^{2}+12(2a-3)\mathbf{c}+4(2a+3)}{2\left(3 \mathbf{c} - 2\right)^3}\,.
\label{sub}
\end{equation}
In the above expression, the coupling $a$ appears as an independent
parameter, but it does affect the value of the scalar (\ref{1-u}) and the
entropy (\ref{1-entropy}) until the relevant order in $\epsilon $. However,
as we depart farther from the horizon, the scalar field is well defined only
if (\ref{sub}) is positive, which is possible (with $\mathbf{c}>\frac{2}{3}$
and $A\epsilon>0$) only when the polynomial in the numerator is positive.
Because the numerator geometrically presents a mostly negative hyperbola in $%
\mathbf{c}$, the conditions in $a$ have to ensure an existence of a positive
section of the hyperbola, that is,
\begin{equation}
\frac{2}{3}<\mathbf{c}\leq \frac{2}{11}\left( 2a-3+\frac{1}{3}\sqrt{a^{2}-%
\frac{7a}{6}+5}\right) \,,
\end{equation}
where one root is smaller than $\frac 23$. The right bound is real for any $%
a $ and it is bigger than $\frac{2}{3}$ when $a>\frac{9}{10}$. Thus, there
are infinitely many possibilities for the coupling constants $a$ and $%
\mathbf{c}$ leading to the scalarization. Interestingly, neither too strong
nor too weak coupling $\mathbf{c}$ will produce a scalarization. This
situation (existence of the solution) is generic in all cases \textbf{A},
\textbf{B} or \textbf{C}, and we will not analyze it in other cases, as it
does not influence the physics in the vicinity of the critical point.

\paragraph{Class B}

Within this class, the extremal black hole without scalar hair is an
electrically charged and asymptotically AdS$_{5}$ Tangherlini black hole.
This scalar field has mass $m^{2}=\frac{6}{5}$, the strength of the
interaction is measured with respect to $\mathbf{c}_0=2\left(\frac
35\right)^4=\frac{162}{625} $, and the effective coupling is $\mathbf{c}>%
\frac{162}{625}\simeq 0,29$. Hence, the entropy density has the form
\begin{equation}
S(Q)=\left\{
\begin{array}{lll}
6\sqrt{3}\pi +\dfrac{6\pi }{5}\,\left( Q-Q_{c}\right) +\dfrac{\sqrt{3}\pi }{%
250}\,\left( Q-Q_{c}\right) ^{2}+\mathcal{\cdots }\,, & Q\leq Q_{c}, & \text{%
Tangherlini~AdS}_{5}~\text{BH;} \\
&  &  \\
6\sqrt{3}\pi +\dfrac{6\pi }{5}\,\left( Q-Q_{c}\right) +\dfrac{\sqrt{3}\pi }{%
250}\dfrac{\left( \mathbf{c}-\frac{6}{25}\right) }{\left( \mathbf{c}-\frac{%
162}{625}\right) }\,\left( Q-Q_{c}\right) ^{2}+\mathcal{\cdots }\,, & Q\geq
Q_{c}, & \text{sph.~hairy~AdS}_{5}~\text{BH,}%
\end{array}
\right.  \label{2-entropy}
\end{equation}
and the order parameter reads
\begin{equation}
u(Q)=\left\{
\begin{array}{ll}
0\,, & Q\leq Q_{c}\,,\medskip \\
\frac{2}{5\sqrt{5}\sqrt[4]{3}}\sqrt{\dfrac{Q-Q_{c}}{\mathbf{c}-\frac{162}{625%
}}}+\cdots \,, & Q\geq Q_{c}.%
\end{array}
\right.
\end{equation}
Thus, for small charges, i.e. for $Q\leq Q_{c}$, the extremal black hole is
an asymptotically AdS$_{5}$ Tangherlini black hole, with no scalar hair.
Since $\dfrac{\mathbf{c}-\frac{6}{25}}{\mathbf{c}-\frac{162}{625}}>1$, as
the electric charge density decreases to approach $Q_{c}$ from the left, the
scalar field starts to condensate on the horizon, because when $Q\geq Q_{c}$
the corresponding hairy extremal black hole has a higher entropy. Again, as
resulting from (\ref{2-entropy}), the phase transition corresponding to the
scalarization is of second order.

\paragraph{Class C (\textit{light} scalar)}

Within this class, the extremal black hole without scalar hair is the
analogue of an electrically charged and asymptotically AdS$_{5}$ Tangherlini
black hole, but with hyperbolic horizon ($k=-1$); let us denote it with
``hypTangh.". Consider a `light' scalar field with mass $m^{2}=\frac{1}{2}$,
with the strength of the effective interaction measured with respect to $%
\mathbf{c}_0=\frac{13}{12}$, and given by $\mathbf{c}>\frac{13}{12}$. Hence,
the entropy density has the form
\begin{equation}
S(Q)=\left\{
\begin{array}{lll}
\sqrt{2}\pi +16\pi \left( Q-Q_{c}\right) +\dfrac{\pi }{48}\,\left(
Q-Q_{c}\right) ^{2}+\mathcal{\cdots }\,, & Q\geq Q_{c}, & \text{%
hypTangh.~AdS }_{5}~\text{BH;} \\
&  &  \\
\sqrt{2}\pi +16\pi \left( Q-Q_{c}\right) +\dfrac{\pi }{48}\dfrac{\left(
\mathbf{c}+\frac{1}{4}\right) }{\left( \mathbf{c}-\frac{13}{12}\right) }%
\,\left( Q-Q_{c}\right) ^{2}+\mathcal{\cdots }\,, & Q\leq Q_{c}, & \text{%
hyp.~hairy~AdS}_{5}~\text{BH,}%
\end{array}
\right.  \label{3-entropy}
\end{equation}
and the order parameter is
\begin{equation}
u(Q)=\left\{
\begin{array}{ll}
0\,, & Q\geq Q_{c}\,,\medskip \\
\frac{\sqrt[4]{2}}{2\sqrt{3}}\sqrt{\dfrac{Q_{c}-Q}{\mathbf{c}-\frac{13}{12}}}
+\cdots \,, & Q\leq Q_{c}\,.%
\end{array}
\right.
\end{equation}
For large charges $Q\geq Q_{c}$, the extremal black hole is an
asymptotically AdS$_{5}$ hyperbolic Tangherlini black hole, with no scalar
hair. Because $\dfrac{\mathbf{c}+\frac{1}{4}}{\mathbf{c}-\frac{13}{12}}>1$,
as the electric charge density decreases to approach $Q_{c}$ from the right,
the scalar field condensates on the horizon since, when $Q\leq Q_{c}$, the
corresponding hairy extremal black hole has a higher entropy. As in the
other cases, the phase transition (\ref{3-entropy}) corresponding to the
scalarization is of second order.

\paragraph{Class C (\textit{heavy }scalar)}

Within the same class, we can consider a `heavy' scalar field with mass $%
m^{2}=\frac{4}{5}$, with the strength of the effective interaction compared
to $\mathbf{c}_0=\frac{91}{3}\left(\frac 25\right)^4=\frac{1,456}{1,875}$
and given by $\mathbf{c}>\frac{1,456}{1,875}$. Hence, the entropy density
has the form%
\begin{eqnarray}
S(Q) &=&\left\{
\begin{array}{lll}
S_{01}(Q) -\dfrac{\sqrt{7}\pi }{5,250}\,\left( Q-Q_{c}\right) ^{2}+\mathcal{%
\cdots }\,, & Q\geq Q_{c}, & \text{hypTangh.~AdS}_{5}~\text{BH;} \\
&  &  \\
S_{01}(Q) -\dfrac{\sqrt{7}\pi }{5,250}\dfrac{\left( \mathbf{c}-\frac{4,200}{%
1,875}\right) }{\left( \mathbf{c}-\frac{1,456}{1,875}\right) }\,\left(
Q-Q_{c}\right) ^{2}+\mathcal{\cdots }\,, & Q\leq Q_{c} & \text{hyp.~hairy~AdS%
}_{5}~\text{BH,}%
\end{array}
\right.  \notag \\
&&  \label{4-entropy}
\end{eqnarray}
where $S_{01}(Q):=14\sqrt{7}\pi +\dfrac{2\sqrt{14}\pi }{5}\,\left(
Q-Q_{c}\right)$ and its derivatives are continuous in the limit $Q \to Q_c$,
and the order parameter is
\begin{equation}
u(Q)=\left\{
\begin{array}{ll}
0\,, & Q\geq Q_{c}\,, \\
\frac{2\sqrt[4]{2}}{5\sqrt{5}\sqrt{3}}\sqrt{\dfrac{Q_{c}-Q}{\mathbf{c}-\frac{%
1,456}{1,875}}}+\mathcal{\cdots }\,, & Q\leq Q_{c}\,,\medskip%
\end{array}
\right.
\end{equation}
Differently from all other classes treated above, in the `heavy scalar'
regime of the class \textbf{C} the second order term in the near-critical
expansion of the extremal black hole entropy becomes negative, and thus a
separate analysis is deserved.

For large charges, i.e. for $Q\geq Q_{c}$, the extremal black hole is an
asymptotically AdS$_{5}$ hyperbolic Tangherlini black hole. Since $\dfrac{%
\mathbf{c}-\frac{4,200}{1,875}}{\mathbf{c}-\frac{1,456}{1,875}}<1$ (and $%
\lim\limits_{\mathbf{c}\rightarrow \infty } \dfrac{\mathbf{c}-\frac{4,200}{%
1,875}}{\mathbf{c}-\frac{1,456}{1,875}}=1^{-}$), as the electric charge
density decreases to approach $Q_{c}$ from the right, the scalar field
condensates on the horizon, because when $Q\leq Q_{c}$ the corresponding
hairy extremal black hole has a higher entropy. From (\ref{4-entropy}), we
can conclude that the phase transition corresponding to the scalarization is
of second order.


\section{Conclusions}
\label{Conclusions}

We discuss thermodynamics of static, electrically charged extremal black holes in Einstein-Maxwell gravity coupled to a complex scalar field in $D$ dimensions, in presence of an arbitrary cosmological constant $\Lambda $. A
non-minimal coupling of the scalar field is described by the non-linear Stueckelberg interaction and the Higgs potential, characterized by the
coupling constants $a$ and $b$, respectively. The former modifies the kinetic term of the scalar field whereas the latter modifies the potential.
When $a,b=0$, the coupling becomes minimal.

The relevant thermodynamic quantity of the extremal black holes is the
entropy. We compute it using the entropy function formalism and analyse its
global maxima describing stable states. We show that there is always a
critical point, with the electric charge density $Q_{c}$, where the scalar
field spontaneously condensates, leading to a thermodynamic instability of
the Reissner-Nordstr\"{o}m black hole. For a given background with $\Lambda
\neq 0$ and the non-planar black hole horizon $k\neq 0$, we find three
regions in the space of the solutions, called \textbf{A}, \textbf{B} and
\textbf{C}, determined by the positive intervals of the scalar mass, $%
\varkappa m^{2}\neq 1,\frac{D-2}{D-3}$. In these sectors, the hairy extremal
black hole is thermodynamically more stable than the electrically charged
extremal Reissner-Nordstr\"{o}m-(A)dS one, when the effective interaction is
strong enough, $\mathbf{c}>\mathbf{c}_{0}$. We determine the critical
exponents in the vicinity of the critical point. In particular, the order
parameter has critical exponent $1/2$, the same as in the Landau-Ginsburg
theory. In the Ehrenfest-like classification based on the behavior of the
entropy, this phase transition is of second order, because the response
function $S^{\prime \prime }(Q)$ has a discontinuity in $Q_{c}$. Out of the
regions \textbf{A}, \textbf{B} and \textbf{C}, the extremal black hole is
thermodynamically stable.

The effective coupling constant $\mathbf{c}=m^{2}a-2b$ appears due to the
attractor mechanism of the extremal black holes, yielding a proper attractor
value of the scalar field at the event horizon, $u$, for a positive
effective mass, $m_{\mathrm{eff}}^{2}(u)=2A\mathbf{c}\left( Q_{c}-Q\right)
>0 $. Since $m_{\mathrm{eff}}^{2}(u)$ and $m_{\mathrm{eff}}^{2}(0)$ have
opposite signs, when $u=0$ is a proper attractor, then $u\neq 0$ is a
repeller, and \textit{vice versa}. Away from criticality ($Q-Q_{c}$ is not
small), both $a$ and $b$ play important roles, governed by the field
equations, such that the Stueckelberg interaction modifies the kinetic term
as in the $\sigma $-model, while the Higgs interaction modifies only the
potential.

We work out five-dimensional case in detail. A typical process of
scalarization goes as follows (on the example of the extremal black hole
belonging to the class \textbf{A}): the strongly coupled ($\mathbf{c}>%
\mathbf{c}_{0}$) Tangherlini dS$_{5}$ extremal black hole, electrically
charged with $Q>Q_{c}$, has vanishing scalar field. When we slowly decrease
the charge so that the black hole remains extremal, after the critical point
($Q<Q_{c}$), the scalar field will start to condensate on the horizon, and
forming a hairy black hole, which has larger entropy. This system describes
a phase transition of second order. Similar situation occurs in cases
\textbf{B} and \textbf{C.}

There are still many open questions related to spontaneous scalarization of
extremal black holes. One of the most important ones is to find an explicit
solution of the Einstein-Maxwell equations for a given boundary conditions,
in one of the sectors \textbf{A}, \textbf{B} or \textbf{C}, which shows a
phase transition. To this end, one should choose a good radial coordinate to
be able to impose correct boundary conditions for extremal black holes, by
introducing a warp factor and the \textit{extremality parameter}, $\xi $,
which is a measure of deviation of two different horizons $r_{+}>r_{-}$ from
their extremal value $r_{h}$, that is, $r_{\pm }=r_{h}\pm \xi$. For the
scalar to be regular, one needs the physical distance $\mathrm{d}\rho ^{2}=%
\mathrm{e}^{-2U}\mathrm{d}r^{2}$ and $\Psi (\rho _{h})$ to be finite. Note
that our radial coordinate appearing in AdS$_{2}$ is the distance from the
horizon (thus located at $r=0$).

Another important step is embedding this model in supergravity and studying
similar instabilities using particular forms of $V(\Psi )$ and $P(\Psi )$
coming from the gauged supergravity. In particular, it would be interesting
to make $\sigma (x)$ a dynamic field within the $\sigma$-model, and study the exponential potential between the scalar field and the electrodynamics, as
typically occurs in supergravity. Concerning the embedding in $D=5$
supergravity, since the model discussed here contains one Maxwell field and
one complex scalar field, it cannot be regarded as the (purely) bosonic
sector of any $\mathcal{N}\geq 2$ gauged supergravity. Still, it might be
embedded in gauged supergravity with suitable truncation of the bosonic
sector (namely, of the scalar and vector fields). If this were possible, it
would then be interesting to see whether the relation between the $Q$, $%
\Lambda $ and $k$ is related to the BPS states. We leave the investigation
of these issues to future work.

We also leave for further study an analysis of asymptotic symmetries near
the horizon in the vicinity of the critical point, looking for the symmetry
enhancement related to the scale invariance. We expect that only the
effective constant $\mathbf{c}$ will play a relevant role in this
approximation.

\section*{Acknowledgments}

The authors would like to thank to Laura Andrianopoli, Antonio Gallerati, Radouane Gannouji and Mario Trigiante for helpful discussions. The work of AM is supported by a \textquotedblleft Maria Zambrano" distinguished researcher fellowship, financed by the European Union within the NextGenerationEU program. The work of OM and PQL was funded in part by FONDECYT Grant N$^{\circ }$1190533, VRIEA-PUCV Grant N$^{\circ }$123.764, and by
ANID-SCIA-ANILLO ACT210100.

\end{document}